\documentclass[fleqn,usenatbib]{mnras}

\usepackage{newtxtext,newtxmath}

\usepackage[T1]{fontenc}
\usepackage{ae,aecompl}

\usepackage{graphicx}
\usepackage{amsmath}

\usepackage{latexdiff}

\title[RRL rotation in the Galactic bulge]{Kinematics of RR Lyrae stars in the Galactic bulge with OGLE-IV and Gaia DR2}

\author[Hangci Du et al.]{Hangci Du,$^{1}$\thanks{E-mail: hangci.du@outlook.com}
Shude Mao,$^{1,2}$
E. Athanassoula,$^{3}$
\newauthor
Juntai Shen,$^{4,5,6,7}$
and Pawel Pietrukowicz$^{8}$ 
\\
$^{1}$Department of Astronomy and Tsinghua Center for Astrophysics, Tsinghua University, 100084 Beijing, China\\
$^{2}$National Astronomical Observatories, Chinese Academy of Sciences, 20A Datun Road, Chaoyang District, Beijing 100101, China\\
$^{3}$Aix Marseille Universit{\'e}, CNRS, CNES, LAM,  Marseille, 13388 Marseille Cedex 13, France\\
$^{4}$Department of Astronomy, School of Physics and Astronomy, Shanghai
Jiao Tong University, 800 Dongchuan Road, Shanghai 200240, China; \\
$^{5}$Shanghai Key Laboratory for Particle Physics and Cosmology, 200240,
Shanghai, China\\
$^{6}$Shanghai Astronomical Observatory, Chinese Academy of Sciences, 80
Nandan Road, Shanghai 200030, China\\
$^{7}$College of Astronomy and Space Sciences, University of Chinese Academy
of Sciences, 19A Yuquan Road, Beijing 100049, China\\
$^{8}$Astronomical Observatory, University of Warsaw, Al. Ujazdowskie 4,00-478 Warszawa, Poland\\
}

\date{Accepted 2020 August 24. Received August 21; in original form 2020 March 16}

\pubyear{2020}

\begin{document}
\label{firstpage}
\pagerange{\pageref{firstpage}--\pageref{lastpage}}
\maketitle

\begin{abstract}
We analyze the kinematics and spatial distribution of 15,599 fundamental-mode RR Lyrae (RRL) stars in the Milky Way bulge by combining OGLE-IV photometric data and Gaia DR2 proper motions. We show that the longitudinal proper motions and the line-of-sight velocities can give similar results for the rotation in the Galactic central regions. The angular velocity of bulge RRLs is found to be around $35$\,km\,s$^{-1}$\,kpc$^{-1}$, significantly smaller than that for the majority of bulge stars ($50-60$\,km\,s$^{-1}$\,kpc$^{-1}$); bulge RRLs have larger velocity dispersion (120$-$140\,km\,s$^{-1}$) than younger stars. The dependence of the kinematics of the bulge RRLs on their metallicities is shown by their rotation curves and spatial distributions. Metal-poor RRLs ([Fe/H]<$-1$) show a smaller bar angle than metal-rich ones. We also find clues suggesting that RRLs in the bulge are not dominated by halo stars. These results might explain some previous conflicting results over bulge RRLs and help understand the chemodynamical evolution of the Galactic bulge.
\end{abstract}


\section{Introduction}
\label{sec:intro}

RR Lyraes (RRL, hereafter) are pulsating, low-metallicity, core-helium-burning horizontal branch giants with age >11 Gyr \citep{1989PASP..101..570W} and trace an old, relatively metal-poor population in the bulge \citep{2018ARA&A..56..223B}. They are standard candles for distance determination using the period-luminosity relation. These properties make RRL population an important tracer to study the bulge dynamics and evolution, even though they are estimated to represent only 1\% of the Galactic bulge population \citep{2012ApJ...750..169P, 2013ApJ...769...88N}.

There have been several disputes over the properties of RRLs in the Galactic bulge \footnote{sometimes referred to as the "inner Galaxy" in the literature}. In terms of the structure of RR Lyraes, \citet{2013ApJ...776L..19D} argued that bulge RRLs, with a bar angle of $12^{\circ}.5\pm 0^\circ.5$, do not trace the standard bar structure (bar angle$\sim27^\circ$,  \citealt{2013MNRAS.435.1874W}); \citet{2016A&A...591A.145G} concluded that RRLs in high Galactic latitude ($-10.3^\circ \lesssim b\lesssim -8.0^\circ$) show no evidence of an X-shaped structure, which also supports the postulate that RR Lyraes do not follow the bar structure; \citet{2015ApJ...811..113P} analyzed the bulge RRL sample from the OGLE-IV survey \citep{2014AcA....64..177S} and concluded that the bar angle is $\sim 21^\circ$, considerably larger than that from \citet{2013ApJ...776L..19D}, they also presented that the RRLs show evidence of a triaxial distribution. In terms of the origin of RR Lyraes, \citet{2018ApJ...857...54D} proposed that bulge RRLs might correspond to the outskirts of an ancient Galactic spheroid or classical bulge component residing in the Galactic centre, while  \citet{2018MNRAS.479..211M} favoured a thick disc with short scale height and short scale length. In terms of the kinematics of RR Lyraes, \citet{2019MNRAS.485.3296W} showed that the rotation of RRLs at a distance of 1.5\,kpc from the Galactic centre is $\sim 50$\,km\,s$^{-1}$, which, as he said, is in disagreement with the model by \citet{2017MNRAS.464L..80P}.

Line-of-sight (LOS, hereafter) velocity and proper motion are two kinds of fundamental kinematics observables, both of which can be used to generate rotation curves in the Galactic bulge (e.g., \citealt{2013ApJ...769...88N, 2018ApJ...858...46C}); our paper will connect these two kinds of rotation curves. Due to the technical progress in massive spectroscopic observations, there have been many LOS velocity surveys probing stars toward the Galactic bulge in the past ten years: BRAVA \citep{2008ApJ...688.1060H,2012AJ....143...57K}, ARGOS \citep{2013MNRAS.428.3660F,2013MNRAS.432.2092N}, GIBS \citep{2017A&A...599A..12Z}, GES \citep{2014A&A...569A.103R}, APOGEE \citep{2016ApJ...832..132Z, 2016ApJ...819....2N}. In contract, proper motion (PM) surveys are much less. HST SWEEPS survey \citep{2008ApJ...684.1110C,2018ApJ...858...46C} tried to use proper motions as bulge rotation indicators. However, it only gave a qualitative but not quantitative description due to their 
large distance uncertainty. In our work, we will demonstrate the equivalence of the proper motion and the line-of-sight velocity as bulge rotation indicators. Thanks to Gaia DR2 \citep{2016A&A...595A...1G,2018A&A...616A..11G} with proper motions and OGLE-IV RR Lyrae sample \citep{2015ApJ...811..113P} with distance, we can study bulge kinematics using proper motions. \citet{2019MNRAS.485.3296W} have dealt with PanSTARRS1 halo RRLs $1.5-20$\,kpc from the Galactic centre, while our work concentrates on RRLs in the innermost 1.5\,kpc from the Galactic centre.

A further aim of this paper is to use RRL stars to probe the metallicity-kinematics correlation in the bulge region and the bulge formation history. Several surveys have studied the kinematics of other types of bulge stars with different metallicities. Within a radius of $3.5$\,kpc around the Galactic centre, \citet{2013MNRAS.432.2092N} found different kinematics in four metallicity bins, among which the stars with [Fe/H]$>-0.5$ show a near-cylindrical, faster rotation, while the stars with [Fe/H]$<-0.5$ have a significantly slower rotation, which is consistent with \citet{2016ApJ...819....2N} results combining APOGEE and ARGOS data. GIBS \citep{2017A&A...599A..12Z} divides their targeting red clumps and red-giant-branch stars in the inner bulge ($|l|\leq 8.5^\circ, b=1.4-8.5^\circ$) into two groups separated approximately at [Fe/H]=0, showing that the metal-poor component has a higher radial velocity dispersion compared to the metal-rich one at all longitudes. GES \citep{2014A&A...569A.103R,2017A&A...601A.140R} gave similar results as GIBS. HST SWEEPS \citep{2018ApJ...858...46C} also found an evident trend that metal-poor bulge stars rotate slower than metal-rich ones. A slower rotation curve was also found for the metal-poor stars in PIGS \citep{2020MNRAS.491L..11A}. This phenomenon also appears in our work.

The paper is organized as follows: In section~\ref{sec:data}, we briefly introduce the sample we use. In section~\ref{sec:comp}, we derive the relationship of PM and LOS velocities as bulge rotation indicators geometrically and validate it with mock data. In section~\ref{sec:result}, we show several results from the RRL distribution in spatial and velocity space and compare them with results from other surveys. In section~\ref{sec:discussion}, we explain why previous surveys on RRLs in the Galactic bulge did not observe the rotation quantitatively. We summarize and conclude in section~\ref{sec:conclusion}. 

\section{Data}
\label{sec:data}

\begin{figure}
 \includegraphics[width=\columnwidth]{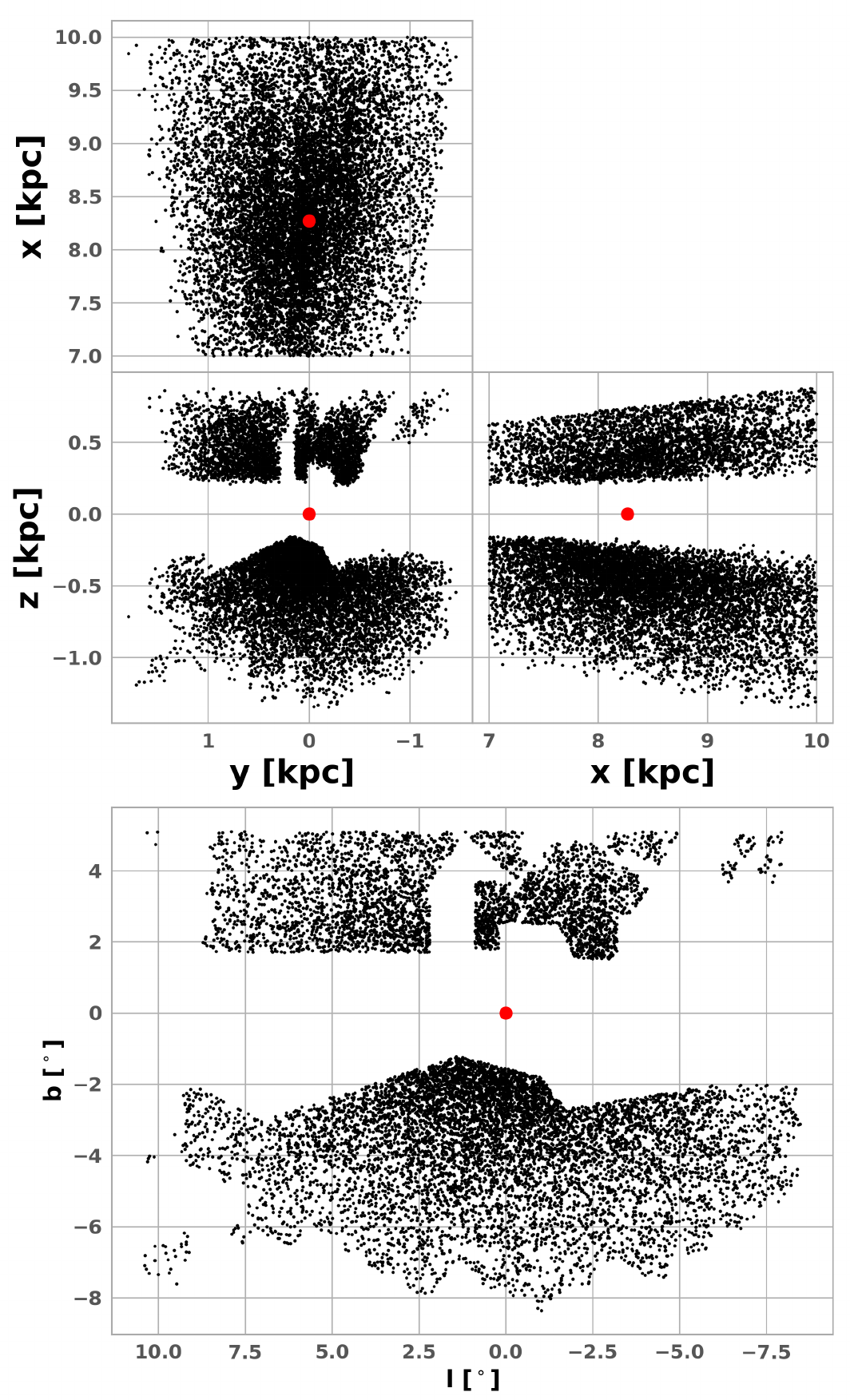}
 \caption{Face-on and side-on distribution of the OGLExGaia sample in Cartesian coordinate system; red dots are the Galactic center. Our whole sample covers $0\sim 20$\,kpc in the x-direction, while we only show the inner 3\,kpc for better visualization. For consistency, we use here the distance to the Galactic centre (GC) of 8.27\,kpc \citep{2015ApJ...811..113P}.}
 \label{fig:space}
\end{figure}

\begin{figure*}
    \includegraphics[width=\linewidth]{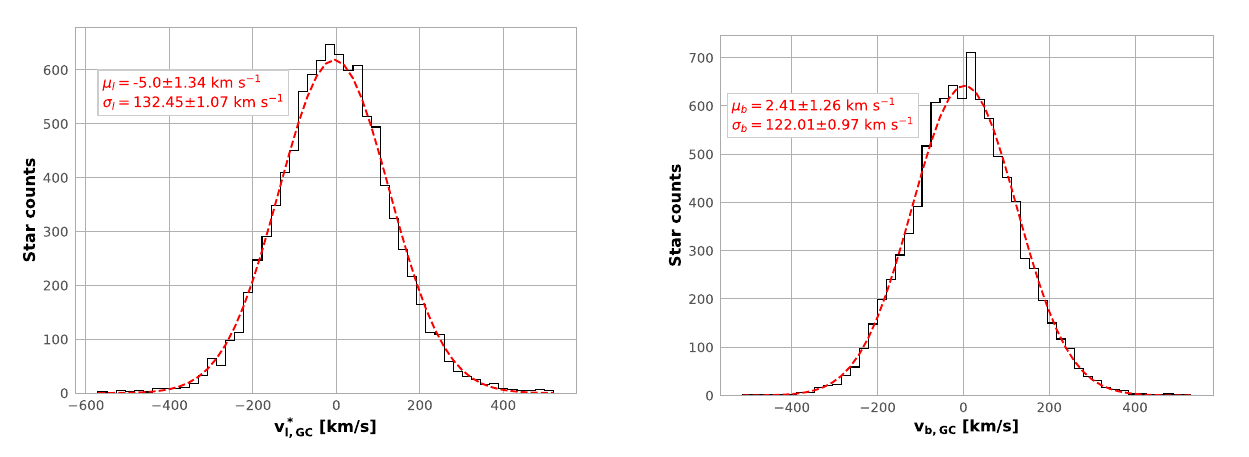}
    \caption{The transverse velocity distributions derived from proper motions of OGLExGaia RRLs in $l$ and $b$. The best Gaussian fit is shown as the dashed curve; the mean and dispersion are indicated in each panel. We can see $\langle v_b \rangle\sim 0$, dispersion ratio $\sigma_l/\sigma_b\sim1.1$; both are consistent with previous results \citep{2004ApJ...616..872R,2019MNRAS.487.5188S,2019MNRAS.489.3519C}. }
  \label{fig:ogle_hist}
\end{figure*}

The original sample of our work is from \citet{2014AcA....64..177S, 2015AcA....65....1U}, which got an update by \citet{2019AcA....69..321S}. The sample consists of 38,257 RRLs over 182 square degrees (see \citealt[their Figure~2]{2014AcA....64..177S} for the coverage on the sky), within which 27,258 are high-amplitude fundamental-mode RRab type stars. The photometry of the catalog was based on the Cousins I-band with 100\,s exposure (ranging from about 13\,mag to 20.5\,mag). The Johnson V-band with 150\,s exposure was also added for color information. RRLs were identified by Fourier analyses of the light curves. Metallicities were obtained from the Fourier coefficients. The completeness of the RRab stars in the sample is estimated to be $96\%\sim 97\%$ \citep{2014AcA....64..177S, 2019AcA....69..321S}, which means we can safely obtain the star-count map and the spatial number density.

We follow the cleaning process by \citet{2015ApJ...811..113P} as follows: we first take the sample of 27,258 RRab variables and reject 54 stars as being bona fide members and very likely members of eight globular clusters. Second, the sample was cleaned from the foreground and background RR Lyrae stars by constructing the color-magnitude (V-I, I) diagram \citep[their Figure~1]{2015ApJ...811..113P}, leading to a sample of 21,026 objects. It was also based on the color-magnitude diagram that the sample of RRab stars is complete down to I=18\,mag. Finally, the region with a mean brightness of I>18\,mag \citep[their Figure~2]{2015ApJ...811..113P} is excluded, leaving with only the "complete" area with all RRab stars presumed to be detected, which amounts to 16221 objects in 90.5~deg$^2$,.

\begin{table}
\caption{Match radius and match rate. "Best" means the closest inside the radius while "All" means all in the radius.}
\label{tab:cross_match}
\begin{tabular}{lllll}
\hline
Match radius ($\arcsec$) & Best  & All   & All-Best & Match rate \\
\hline
0.4             & 16069 & 16090 & 21       & 99.06\%      \\
0.3             & 16027 & 16032 & 5        & 98.80\%       \\
0.2             & 15599 & 15600 & 1        & 96.17\%       \\
0.18            & 15218 & 15218 & 0        & 93.82\%       \\
0.16            & 14630 & 14630 & 0        & 90.19\%       \\
\hline
\end{tabular}
\end{table}

The distance of the RRab stars is determined according to \cite{2015ApJ...811..113P}:
\begin{equation}
d=10^{1+0.2(I_0-M_I)}=10^{1+0.2(I-A_I-M_I)},
\end{equation}
\noindent where $A_I$ was derived from the formula introduced in \citet{2013ApJ...769...88N}:
\begin{equation}
A_{I}=0.7465 E(V-I)+1.3700 E(J-K).
\end{equation}
\noindent The reddening in the optical regime is $E(V-I)=(V-I)-(V-I)_0=(V-I)-(M_V-M_I)$. The absolute brightnesses $M_V$ and $M_I$ are computed from the theoretical relations given in \citet{2004ApJS..154..633C}:
\begin{equation}
  \left\{
    \begin{aligned}
    M_{V}&=2.288+0.882 \log Z+0.108(\log Z)^{2},\\
    M_{I}&=0.471-1.132 \log P+0.205 \log Z,\\
    \end{aligned}
  \right.
  \label{eq:PLZ}
\end{equation}
\noindent with the following conversion for metallicity:
\begin{equation}
    \log Z={\rm [Fe/H]}-1.765,
\end{equation}
\noindent where [Fe/H] is defined in Equation~\ref{eq:met}. We also use the RRL period-luminosity-metallicity (PLZ) relationship from \citet{2015ApJ...808...50M} to build another distance dataset for cross-check:
\begin{equation}
    \begin{aligned}
    M_{I}&=-0.07-1.66\log{P}+0.17{\rm [Fe/H]},\\
    \end{aligned}
\end{equation}
\noindent so that we make our results more robust. The reddening $E(J-K)$ was taken from the maps in \citet{2012A&A...543A..13G} which were prepared using the VVV survey \citep{2010NewA...15..433M}. We note that, the VVV calibration issue pointed out by \citet{2020ExA....49..217H} does not influence the extinction map. According to \citet{2011A&A...534A...3G}, the CASU photometry they started from was re-calibrated using 2MASS, which is different from the "standard photometric source catalogues from VVV that comes from VDFS" as from \citet{2020ExA....49..217H}. The comparison of the reddening maps presented in \citet{2012A&A...543A..13G} is furthermore fully consistent with the most recent PSF photometry from \citet{2019A&A...629A...1S}, which was calibrated independently.

Typical statistical error of metallicities and distances are $\sim$0.01 dex and 0.15 kpc for the whole sample, among which the (statistical) photometric metallicity error calculated as d[Fe/H] = 0.824 d$\phi_{31}$ is the derivative of the formula from \citet{2005AcA....55...59S}: 
\begin{equation}
  {\rm [Fe/H]} = -3.142 - 4.902 P + 0.824 \phi_{31},
  \label{eq:met}
\end{equation}
\noindent where $\phi_{31}$ is a combination of Fourier parameters $\phi_{31} = \phi_3 - 3 \phi_1$. According to \citet{2005AcA....55...59S}: the metallicity error is calculated as d[Fe/H] = $0.824 {\rm d}\phi_{31}$; the systematic error of the method is about 0.18\,dex. The distance error is derived from the error propagation, where the error of E(V-I) and E(J-K) is provided by \citet{2013ApJ...769...88N} and \citet{2012A&A...543A..13G} respectively. 



We cross-match the cleaned sample with Gaia DR2, obtaining a sample of 15,599 sources within a $0.2\arcsec$ match radius. We use the CDS cross-match service, within which the positions are propagated from epoch J2015.5 to J2000 when proper motions are available\footnote{\url{https://www.cosmos.esa.int/documents/29201/1773953/Gaia+DR2+primer+version+1.3.pdf/a4459741-6732-7a98-1406-a1bea243df79}}, to match our OGLE-IV catalog and Gaia data. The match radius is determined as follows: as the OGLE-IV median seeing is $\sim 1\arcsec$ \citep{2015AcA....65....1U}, the over-match problem is from the Gaia survey to the crowded bulge area. Table~\ref{tab:cross_match} shows the match result with different match radii and match modes provided by CDS, "best" means the closest inside the match radius while "all" means all in the match radius. We see 0.2$\arcsec$ is where the over-match problem fades away in our field. Overall, the match rate is $\sim$96\%, which is sufficiently high for us to ignore potential selection bias. 

To ensure the Gaia proper motions are reliable, we use the recommended astrometric quality parameter for Gaia DR2\footnote{\url{https://www.cosmos.esa.int/web/gaia/dr2-known-issues}}, the re-normalized unit weight error (RUWE) $\mu$, described by \citet{2018A&A...616A...2L}. A high value of $\mu$ may be caused by partially resolved or astrometric binaries. We select sources with $\mu<1.4$ in our sample \citep{2018A&A...616A...2L}, leading to a sample of 12,337 objects. After this cleaning, we use error propagation from  distances and observed proper motions, to obtain the typical error of $v_l^*$ and $v_b$ (as defined in section~\ref{sec:geo_def}) $\sim$3.5\% and 2.2\% respectively.

We adopt the Cartesian coordinate system as described in section~\ref{sec:geo_def} to our sample, which covers $0\sim20$\,kpc in the $x$-axis (toward the Galactic centre). The distribution of the cleaned sample is shown in Figure~\ref{fig:space} (we only show the 7\,kpc<$x$<10\,kpc part for better visualization, this inner 3\,kpc range is also used for more specific research in the bulge region). We use the distance to the Galactic centre determined by \citet{2015ApJ...811..113P} (8.27\,kpc) for consistency. Figure~\ref{fig:ogle_hist} shows the histogram of the transverse velocities derived from PMs for the whole sample together with two Gaussian best fits. We see the average of $v_b\sim 0$, consistent with \citet{2004ApJ...616..872R}, and the velocity dispersion ratio $\sigma_l/\sigma_b\sim1.1$, consistent with \citet{2019MNRAS.487.5188S}. 


\begin{table*}
\caption{Other surveys for comparison. Now the other catalogs can also be cross-matched with Gaia to generate proper motions. Abbreviations: Multi, multiple populations; RCG, red clump giant; MS, main-sequence object.}
\begin{tabular}{llllll}
\hline
{Project} & {References} & {Population} & {Observable} & {Distance} & {Indicator}  \\\hline
OGLExGaia & Our data set & RRLs & PM & yes & ($-v_l^*,d$) \\\hline
BRAVA & \citet{2008ApJ...688.1060H,2012AJ....143...57K} & M giants & LOS velocity & no & ($v_{\rm  los}^*,R_\perp$)\\
BRAVA-RR & \citet{2016ApJ...821L..25K} & RRLs & LOS velocity & no & ($v_{\rm  los}^*,R_\perp$)\\
ARGOS & \citet{2013MNRAS.432.2092N} & Multi & LOS velocity & no & ($v_{\rm  los}^*,R_\perp$)\\
APOGEE & \citet{2016ApJ...832..132Z,2016ApJ...819....2N} & Multi & LOS velocity & no & ($v_{\rm  los}^*,R_\perp$) \\
GIBS & Zoccali et al. (2017) & RCGs & LOS velocity & no & ($v_{\rm  los}^*,R_\perp$) \\\hline
VVVxGaia & \citet{2019MNRAS.487.5188S} & RCGs & PM & yes & ($-v_l^*,d$) \\
HST SWEEPS & \citet{2008ApJ...684.1110C,2018ApJ...858...46C} & MS & PM & yes & ($-v_l^*,d$) \\\hline
\end{tabular}
\label{tab:surveys}
\end{table*}

Other comparison data sets can be found in Table~\ref{tab:surveys}. We note that nowadays these catalogs can all be cross-matched with Gaia to obtain proper motions.

\section{Comparison of PM and LOS velocities as rotation indicators}
\label{sec:comp}

PM and LOS velocities are both kinematics indicators. Here we will show that the results given by these two indicators are in agreement with each other to show the bulge rotation. We will first use a geometric derivation based on simplest assumptions to give us insights and then validate it with N-body simulations in which the orbits have all the necessary complexity due to the bar. As we will see in Section~\ref{sec:comp_rot}, the results from real data also show consistency (see section~\ref{sec:comp_rot}).

\subsection{Geometrical derivation}
\label{sec:geo}

\begin{figure}
 \includegraphics[width=\columnwidth]{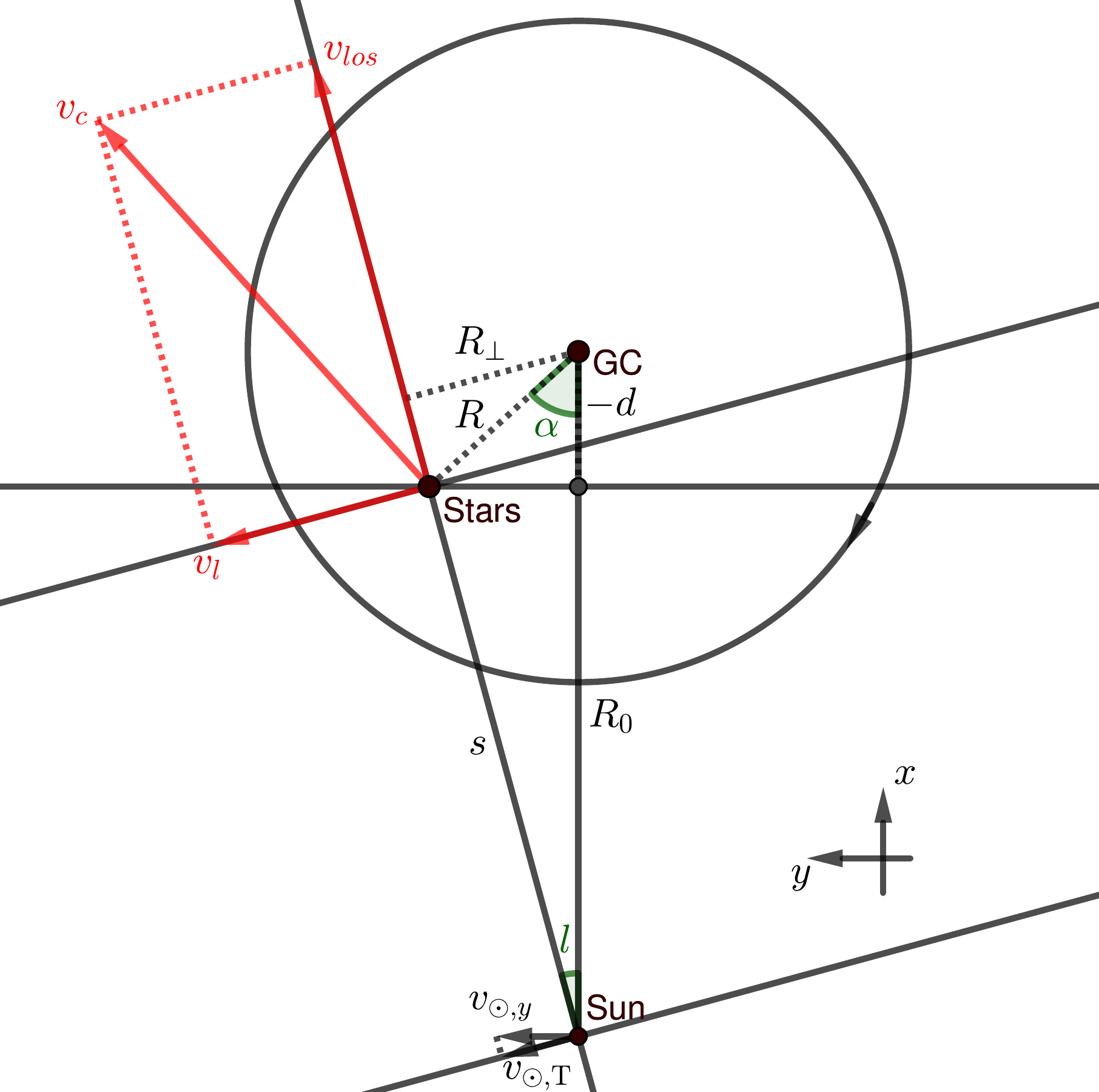}
 \caption{The top-down view from the North Galactic Pole. The red arrows show the circular velocity and its decomposition. The Sun and stars rotate clockwise around the Galactic centre from this view. We define $d\equiv x-R_0, R_\perp\equiv R_0\sin{l}$.}
 \label{fig:geo}
\end{figure}

\subsubsection{Definitions}
\label{sec:geo_def}

We set the Sun as the origin of the Cartesian coordinate system, as shown in Figure~\ref{fig:geo}. The $x$-axis points to the Galactic centre, the $y$-axis points to the positive direction of Galactic longitude, while the $z$-axis points to the North Galactic Pole. From the top-down view, the Sun and stars rotate clockwise around the Galactic centre. 

Here we discuss a star sample in a small region around the Galactic centre in the bulge. We label its velocity as $v_c$, which can be decomposed into PM and LOS velocity. We label the Sun-GC distance as $R_0$, the distance from the Galactic centre to the line of sight as $R_\perp\equiv R_0\sin{l}$, the Sun-stars distance as $s$, the stars' distance to the Galactic centre as $R$. We define $d\equiv x-R_0$ (we add negative signs in Figure~\ref{fig:geo} to make sure all values shown are positive). Other defined angles are labelled in Figure~\ref{fig:geo}. We may simply use $\frac{v_{l}}{\rm km/s} \equiv 4.74\cdot\frac{\mu_{l}}{\rm mas/yr}\cdot \frac{s}{\rm kpc}$ to obtain transverse velocities $v_{l}$ from longitudinal proper motions $\mu_{l}$.

When dealing with real data, we transform from the Heliocentric (HC) coordinate to the Galactocentric (GC) coordinate to subtract the solar motion: 
\begin{equation}
  \left\{
    \begin{aligned}
      v_{\rm los, GC}&= v_{\rm los, HC} +U_\odot \cos{l}\cos{b} 
      \\&\qquad+(V_\odot+V_{\rm LSR})\sin{l}\cos{b} +W_\odot\sin{b},\\
      v_{l, {\rm GC}}^*&=v_{l, {\rm HC}}^* -U_\odot \sin{l}\cos{b}+(V_\odot+V_{\rm LSR})\cos{l}\cos{b},\\
      v_{b, {\rm GC}}&=v_{b, {\rm HC}}+W_\odot\cos{b},\\ 
    \end{aligned}
  \right.
  \label{eq:hc2gc}
\end{equation}

\noindent where $(U_\odot, V_\odot, W_\odot)$ is the solar peculiar velocity relative to the Local Standard of Rest (LSR). We use the Galactocentric velocity as default.

 We note that the dynamics of the inner 5 kpc of the Milky Way disc are strongly influenced by the presence of the Milky Way bar \citep{2019ApJ...871..120E}, which cannot be simply understood as circular motion, but we can use the simplified case to gain insights.
 
\subsubsection{Discussion about the simplest case}
\label{sec:geo_dis}

Temporarily ignoring other complexities, here we first use the simplest model (circular, rigid-body motion) to show that PM and LOS velocities can be equivalently used to determine the bulge rotation, then we use mock and real data to demonstrate that the equivalence is still applicable in the real world, where the simplest model is not necessarily valid.

We assume that the system is in equilibrium and that the velocities are near-circular and rigid-body-like. Based on these assumptions, we know that the rotation curves in the bulge region increase linearly with distance to the Galactic centre, which is not a bad approximation according to the rotation curves from previous surveys \citep{2012AJ....143...57K,2013MNRAS.432.2092N,2019MNRAS.487.5188S}. 

In the Galactic plane, the circular velocity for a star is $v_{c}=\omega R$, where $\omega$ is the angular velocity for the rigid-body rotation, $R$ is its distance from the Galactic centre. Next we will obtain the transverse velocity $v_{l}$ and line-of-sight velocity $v_{\rm los}$ (see Figure~\ref{fig:geo}). For $v_{\rm los}$, we obtain:
\begin{equation}
    v_{\rm los} =\omega R\sin{(\alpha+l)}=\omega R_{\perp}.
    \label{eq:v_los}
\end{equation}
\noindent For $v_{l}$, we know:
\begin{equation}
\begin{aligned}
    v_l &= v_c \cos{(\alpha+l)}\\ &= \omega R \cos{\alpha}(\cos{l}-\tan{\alpha}\sin{l})\\&=\omega(-d)(\cos{l}-\tan{\alpha}\sin{l}).
    \label{eq:vl_01}
\end{aligned}
\end{equation}
\noindent What is more, from the geometry, we know:
\begin{equation}
    (-d)\tan{\alpha} = (R_0+d)\tan{l}.
    \label{eq:vl_02}
\end{equation}
By combining Equation~\ref{eq:vl_01} and Equation~\ref{eq:vl_02}, we obtain:
\begin{equation}
    v_l = \omega[(-d)\cos{l}\cdot(1+\tan^2{l})-R_0\tan{l}\sin{l}].
    \label{eq:v_l}
\end{equation}
Then we have:
\begin{equation}
  \left\{
    \begin{aligned}
      \frac{\partial v_{\rm los}}{\partial R_\perp} &= \omega,\\
      \frac{\partial v_{l}}{\partial(-d)} & =\omega\cos{l}\cdot(1+\tan^2{l})\approx\omega.\\ 
    \end{aligned}
  \right.
  \label{eq:rot_slop}
\end{equation}
The slopes in two diagrams show the same physical property.

\subsection{Validation with mock data}
\label{sec:sim}


\subsubsection{Two simulations}
\label{sec:2sim}

We will now test the consistency between $(-v_l^*,d)$ and $(v_{\rm los}^*,R_\perp)$ using mock data from two quite different simulations using different evolutionary histories.

The first is a grid-based N-body simulation by \citet[the Shen model hereafter]{2010ApJ...720L..72S}. For more details about the mock data, see \citet{2012ApJ...757L...7L}. In the Shen model, 982,889 particles are initially in an unbarred, thin disk. A bar structure emerges during evolution. Then from a bunch of N-body models, they found the one that best matches the BRAVA \citep{2008ApJ...688.1060H} kinematic data after suitable mass scaling. The barred disk evolved from a thin exponential disk that contains $M_{d}=4.25\times10^{10} M_\odot$, about 55\% of the total mass at the truncation radius (5 scale lengths). The scale length and scale height of the initial disk are $\sim$1.9\,kpc and 0.2\,kpc, and the length unit of the simulation is this scale length. The classical bulge in this model is less than 8\%. The model is consistent with data from several surveys \citep{2012AJ....143...57K, 2013MNRAS.435.1874W,2013MNRAS.432.2092N,2016ApJ...832..132Z}. 

We also use a second simulation \citep{2017MNRAS.467L..46A}, as different as possible from the previous one. Contrary to the first simulation described above, this one ($N$-body+SPH simulation of a barred spiral galaxy as a merger remnant) includes a gaseous component and its physics. Note also that all components, including the dark matter halo, are described self-consistently, and that the initial conditions do not include a disc in equilibrium, but the disc is formed during the simulation and its properties evolve with time. We had at our disposal a large number of such simulations, in all of which a massive merger occurred about 8$\sim$10~Gyr ago (e.g., \citealt{2016ApJ...821...90A,2017MNRAS.468..994P}). We chose simulation mdf732 which has already been used in two different studies \citep{2016ApJ...821...90A,2017MNRAS.467L..46A}. It comprises several stellar components with different properties--i.e. a boxy/peanut bulge, thin and thick disc components, and, to lesser extents, a disky pseudo-bulge, a stellar halo and a small classical bulge--all cohabiting in dynamical equilibrium. It has a classical bulge with only 9$-$12 per cent of the total stellar mass and a bar of roughly the correct size, with a boxy/peanut inner part. The mass of each baryonic particle is $10^4 M_{\odot}$, and that of the dark matter ones is $4\times10^4 M_{\odot}$, with 10 and 17.5 million particles in each of these components respectively. The snapshot of mdf732 we use consists of 1,052,821 stellar particles.

\subsubsection{Mock data setup and reduction}

We prepare the mock data from the two simulations with state-of-the-art parameters: we use a Sun-GC distance $R_0=8.178$\,kpc \citep{2019A&A...625L..10G}, a bar angle $\alpha\sim 27^\circ$\citep{2013MNRAS.434..595C,2013MNRAS.435.1874W}, and the solar near-circular motion with velocity $\sim 248$\,km\,s$^{-1}$\citep{2016ARA&A..54..529B}. We use the same Cartesian coordinate system as in section~\ref{sec:geo_def}, where the $x$-axis points towards the Galactic centre. 

To be comparable with real data, we select only simulation particles in the region ($-10^\circ<l<10^\circ, -7^\circ<b<-2.5^\circ, x>0$, where $x>0$ means the line of sight is toward the Galactic centre but not the anti-centre) including Baade's Window, leading to a sample of 47,683 particles (the Shen model) and 37,342 particles (mdf732) respectively.

\subsubsection{Comparison of the two rotation indicators}

\begin{figure*}
 \includegraphics[width=\linewidth]{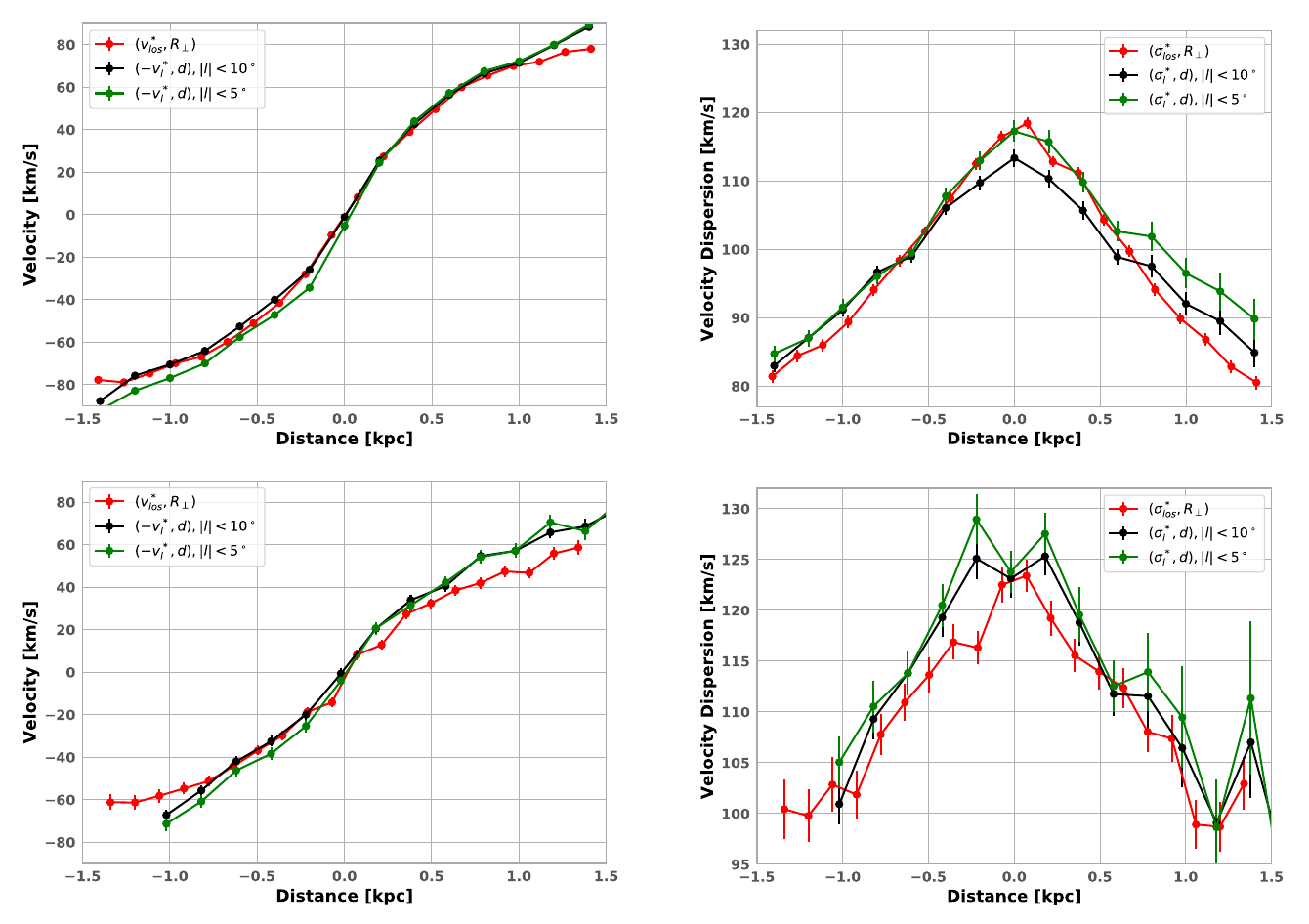}
 \caption{Comparison of two bulge rotation indicators ($v_{\rm los}^*,R_\perp$) (red) and ($-v_l^*,d$) (green and black) based on the mock data of two simulations. The 'Distance' as $x$-axis is $R_\perp$ or $d$ which are defined in Figure~\ref{fig:geo}, the 'Velocity' as $y$-axis is $v_{\rm los}^*$ or ($-v_l^*$). Top panels present results from the Shen model, while bottom panels are from mdf732. Here we use different cut in $l$ to visualize the influence of disc contamination on the results.}
 \label{fig:mock_comp}
\end{figure*}

We compare these two indicators in Figure~\ref{fig:mock_comp}. We label both $R_\perp$ and $d$ as 'Distance' for the $x$-axis, and label both $v_{\rm los}$ and $(-v_l^*)$ as 'Velocity' for the $y$-axis. We see both mock datasets show the consistency between two indicators. If we use a different Galactic longitude cut (i.e. different disc contamination), the consistency still holds.

\subsection{Results with real data}
\label{sec:vali_real}

In Table~\ref{tab:surveys}, we list some kinematic surveys of the Galactic bulge stars. None of these surveys uses both indicators, because we can rarely obtain ($-v_l^*,d$). Before Gaia DR2, PM surveys to the bulge were rare, and it is always hard to obtain the distance from bulge stars. \citet{2018ApJ...858...46C} tried to use PM to describe the rotation of stars in the bulge. However, their accuracy of distance is too poor to give a quantitative description. Thanks to Gaia DR2 and the distance determination of VVV RCGs, the equivalence can be validated by the consistency of the green line and blue \& red lines as shown in section~\ref{sec:comp_rot}, where they describe similar populations and show consistency using different indicators.

\section{Results}
\label{sec:result}

In this section, we quantitatively measure the angular velocity of the RRLs as a distinct slow-rotating, kinematically-hot population in the Galactic bulge, and show solid evidence of the multi-component nature of RRLs, which might explain some previous disputes. 

In this paper, we use the term 'angular velocity' to describe the slope of the rotation curve, which is not the same as the bar pattern speed. The angular velocities are obtained in the inner 1.5\,kpc with linear fitting to the {\it Velocity-Distance} diagrams: for the data points ranging $-1.5{\,\rm kpc}\sim1.5$\,kpc in the $x$-axis, we use the least-squares method to find the best fit; the errors are obtained by the square root of the diagonal elements of the covariance matrix.

In particular, the errors of distances, [Fe/H] and PM are applied with Gaussian resampling as follows: for each value, we create a Gaussian distribution whose mean and error are as observed. For the rotation curves, the error bars are generated with bootstrap with 100,000 samples; for each (re-)sample, we use different, newly-created distances for all stars obeying Gaussian distributions. For the star-count map, we replot it ten more times with newly-created distances to check the influence of distance uncertainty on the bar angle. We use similar method applying the error of [Fe/H] and PM.

\subsection{Cylindrical rotation}
\label{sec:cylin}

\begin{figure*}
 \includegraphics[width=\linewidth]{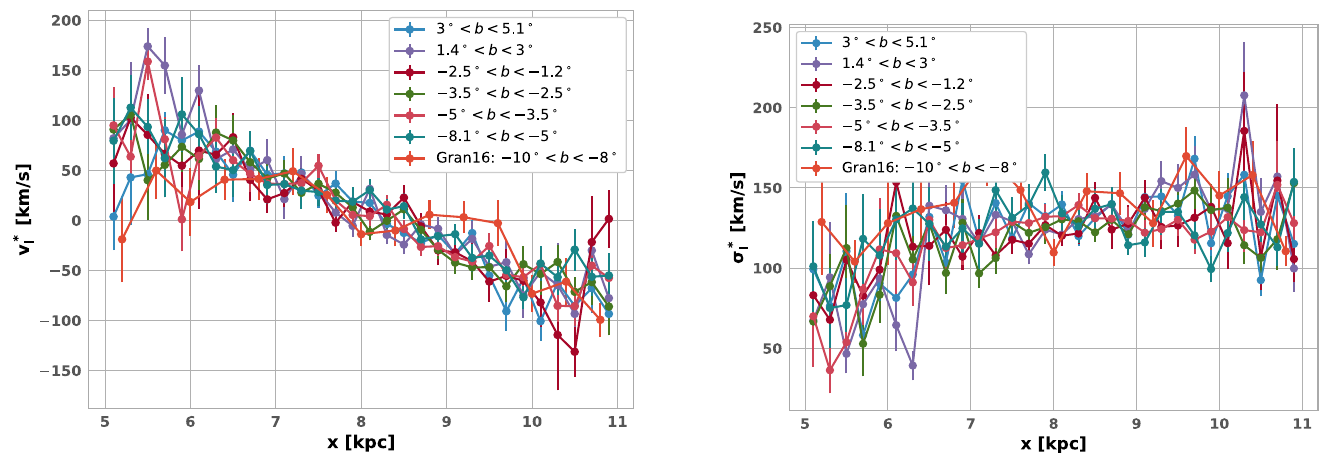}
 \caption{Left panel: Cylindrical rotation of RRLs in different Galactic latitude bins. We divide OGLExGaia data into six bins, including both North and South of the Galactic plane. We also add the sample from a VVV extension program \citep{2016A&A...591A.145G} as completion in high Galactic latitude (labelled as Gran16). The similarity of rotation curves of RRLs in different Galactic latitudes argues for a cylindrical rotation nature, which is similar to the findings from previous surveys on younger populations using LOS velocities (e.g., \citealt{2012AJ....143...57K, 2013MNRAS.432.2092N, 2016ApJ...832..132Z,2017A&A...599A..12Z}). Right panel: Velocity dispersion for the same samples. We also find the absence of the central peak of the velocity dispersion diagram near the Galactic plane ($|b|<5^\circ$), which is similar to metal-poor RCGs \citep[their Figure~6]{2013MNRAS.432.2092N}.}
 \label{fig:cylindrical}
\end{figure*}



We compare the rotation curves at different Galactic latitudes in the left panel of Figure~\ref{fig:cylindrical}. We also add the catalog from \citet[cross-matched with Gaia DR2]{2016A&A...591A.145G} which is a VVV extension program on high Galactic-latitude bulge RRLs ($-10.3^{\circ} \lesssim b \lesssim-8.0^{\circ}, -10.0^\circ\lesssim l\lesssim +10.7^\circ$). To determine the distances, \citet{2016A&A...591A.145G} used the (adapted) period-luminosity-metallicity (PLZ) relation \citep{2015AJ....149...99A}, which modified the PLZ relationship of \citet{2004ApJS..154..633C} to the VIRCAM/VISTA filter system; they applied extinction law from \citet{1989ApJ...345..245C}. We note that their determined-distances might be more significantly influenced by VVV photometric calibration issue \citep{2020ExA....49..217H}.

\begin{figure*}
 \includegraphics[width=\textwidth]{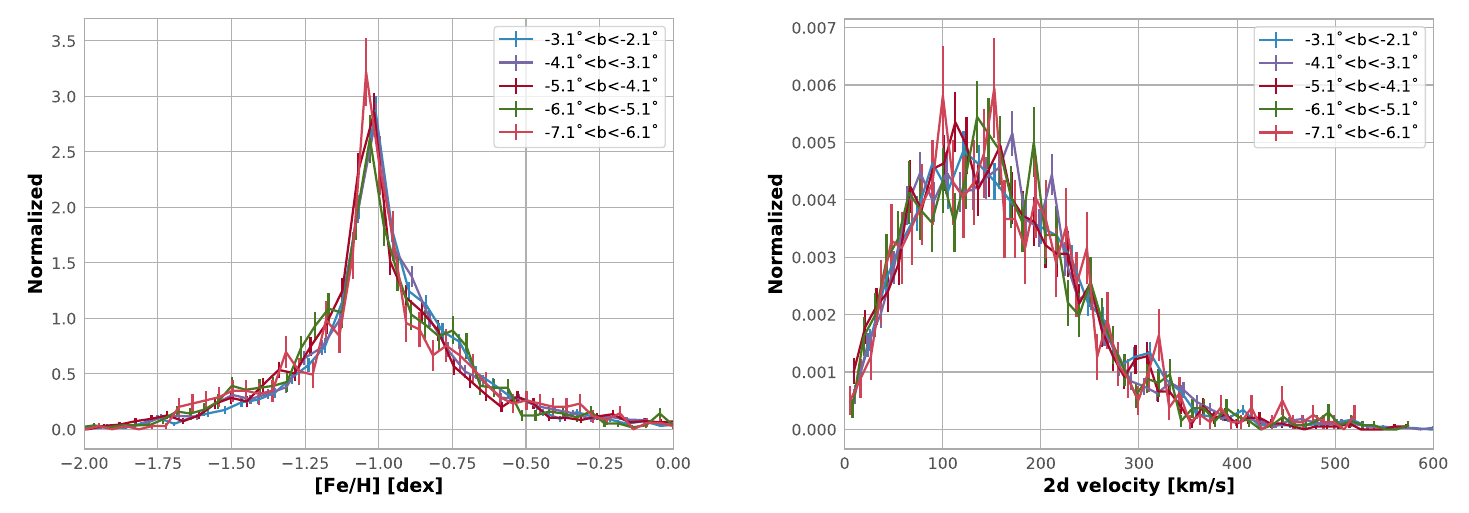}
 \caption{Left: metallicity distribution in different Galactic latitudes. Right: 2D velocity (we define as $\sqrt{(v_l^*)^2+v_b^2}$) distribution in different Galactic latitudes. We see the change of the metallicity and the 2D velocity is mild among different latitudes in our fields. The error bars are Poisson noise.}
 \label{fig:z_distri}
\end{figure*}


Figure~\ref{fig:z_distri} shows the metallicity and 2D velocity (we define as $\sqrt{(v_l^*)^2+v_b^2}$) distributions in different Galactic latitudes. We see the change of the metallicity and the 2D velocity is mild among different latitudes in our fields. From the velocity distribution shown in the right panel of Figure~\ref{fig:z_distri}, we identify high-velocity stars by requiring them to be $>2.5\sigma_{\rm 2D}$. We find this fraction is about 2\%, roughly a factor of 3.5 smaller than the high-velocity star fraction identified in 3D by \citet{2020AJ....159..270K}. We are not sure whether this can be well explained by the difference between 2D and 3D velocities. We return to this issue briefly in the 
conclusion section.

One associated question is whether the slow-rotating and kinematically-hot nature of bulge RRLs is totally caused by the high-velocity stars. We checked the results with and without them (i.e., the results shown in Figures~\ref{fig:surveys} cyan lines), to find this effect do not compensate the high-velocity-dispersion nature; also, the trend of the rotation curve do not change with such cleaning.

Many previous surveys like BRAVA, ARGOS, APOGEE, GIBS have shown the cylindrical rotation nature of younger stellar populations using LOS velocities. This, however, is the first time that the cylindrical rotation of the Galactic bulge is revealed from PMs. 


An interesting fact worth mentioning in the right panel of Figure~\ref{fig:cylindrical} is that there is no central peak, not even for $|b|<3^\circ$, which is not the case of younger populations (for example, M-type Giants in \citealt[their Figure~11]{2012AJ....143...57K}). This is similar to, but more pronounced than what is found for the metal-poor RCGs \citep[their Figure~6]{2013MNRAS.432.2092N}.

\subsection{Rotation of old sub-populations}
\label{sec:multi_met}

\begin{figure*}
 \includegraphics[width=\linewidth]{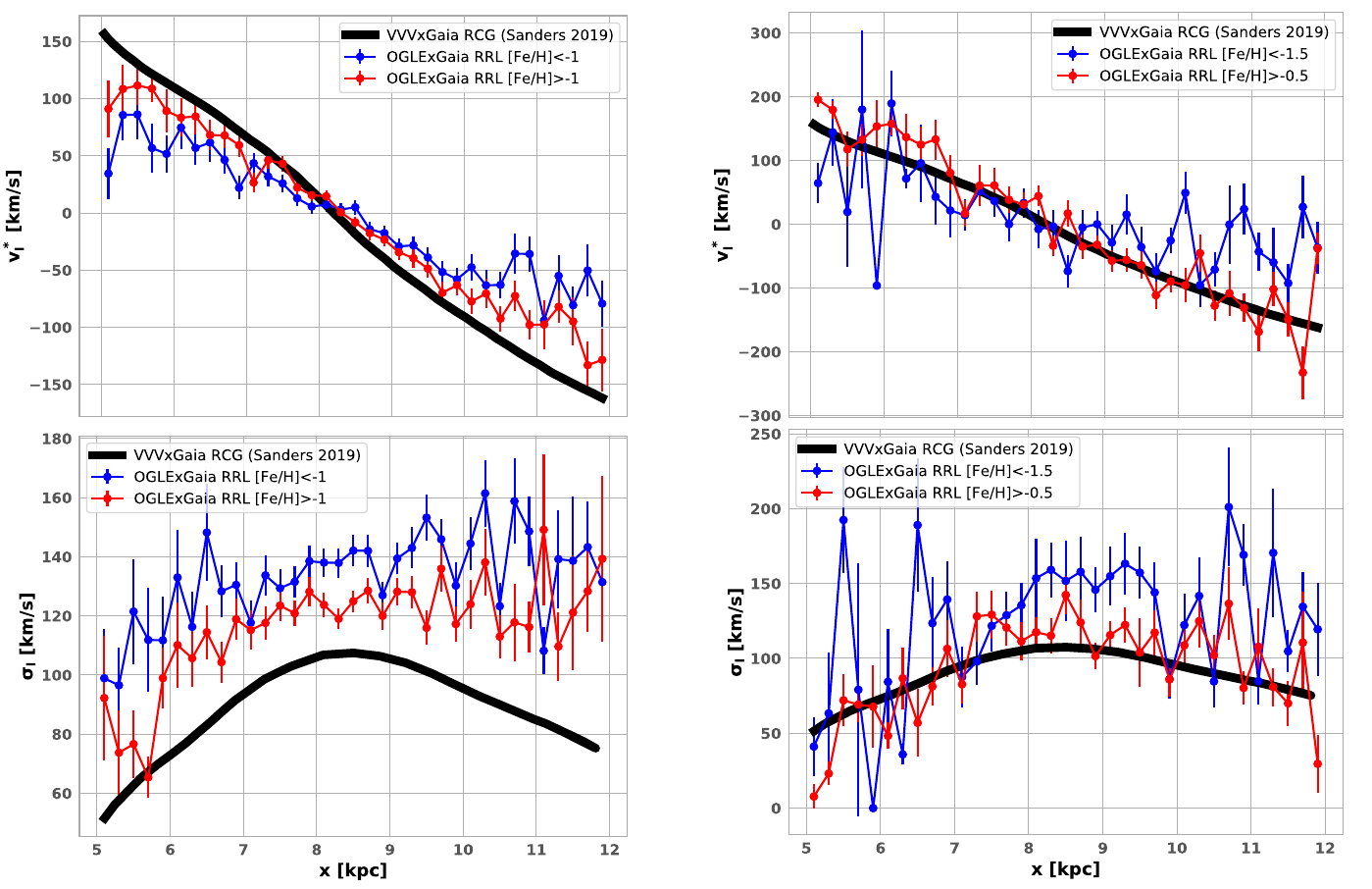}
 \caption{Left panel: rotation (above) and velocity dispersion (below) of metal-poor (blue, [Fe/H]<$-1$) and metal-rich (red, [Fe/H]>$-1$) RRLs. We also add the corresponding curve for  RCGs from \citet{2019MNRAS.487.5188S} for comparison. All the rotation curves are based on the PM from Gaia DR2. The angular velocity of metal-poor RRLs ([Fe/H]<$-1$), metal-rich RRLs ([Fe/H]>$-1$) and RCGs are respectively $(32.42\pm 1.48)$, $(40.29\pm 2.28)$ and $(55.07\pm 0.63)$\,km\,s$^{-1}$\,kpc$^{-1}$. Right panel: as in the left panels, but we now split the sample with stricter restrictions on metallicity. We see RRLs with [Fe/H]$>-0.5$ show properties similar to RCGs. The angular velocity for RRLs with [Fe/H]<$-1$.5 and [Fe/H]>$-0.5$ are respectively $(26.77\pm8.39)$\,km\,s$^{-1}$\,kpc$^{-1}$, $(61.76\pm6.26)$\,km\,s$^{-1}$\,kpc$^{-1}$.}
 \label{fig:rot_diff_met}
\end{figure*}

The multi-component nature of bulge RRLs has been raised recently by \citet[they found two distinct sequences of RRab variables in the period-amplitude diagram.]{2015ApJ...811..113P} and \citet[they found the metallicity distribution function of RRab variables can be fit with multiple Gaussian distributions.]{2018ApJ...857...54D}.

Here we use the metallicity provided by \citet{2015ApJ...811..113P}. \citet[their Figure 17]{2015ApJ...811..113P} shows a sharp and symmetric distribution of RRLs around [Fe/H]$\sim -1$. We thus split our sample in two and obtain the rotation curves shown in the left panel of  Figure~\ref{fig:rot_diff_met}. We also add the result from \citet[their Figure 4]{2019MNRAS.487.5188S} who also used PM data from Gaia DR2.

We see that the metal-poor RRLs rotate the slowest and have the largest velocity dispersion. RCGs, as the main part of the Galactic bulge \citep{2013MNRAS.430..836N}, rotate the fastest and are the most kinematically cold. It seems the metal-rich RRLs are intermediate between metal-poor RRLs and RCGs. Quantitatively, the angular velocity of metal-poor RRLs ([Fe/H]<$-1$), metal-rich RRLs ([Fe/H]>$-1$) and RCGs are respectively $(32.42\pm 1.48)$, $(40.29\pm 2.28)$ and $(55.07\pm 0.63)$\,km\,s$^{-1}$\,kpc$^{-1}$.

We next split the sample again by metallicity, but now considering separately the RRLs with [Fe/H]$<-1.5$ and [Fe/H]$>-0.5$ shown in the right panel. We see that the latter group of RRLs show similar properties to RCGs.

We note that the fact that metal-poorer stars in the Galactic bulge rotate slower and are kinematically-hotter has been observed by several previous bulge surveys for other populations \citep{2013MNRAS.432.2092N, 2016ApJ...819....2N,2017A&A...599A..12Z,2017A&A...601A.140R,2018ApJ...858...46C,2020MNRAS.491L..11A}. This has also been reproduced by N-body chemo-dynamic simulations \citep{2017MNRAS.467L..46A}.

\subsection{Spatial distribution}
\label{sec:spatial}

\begin{figure*}
 \includegraphics[width=0.8\linewidth]{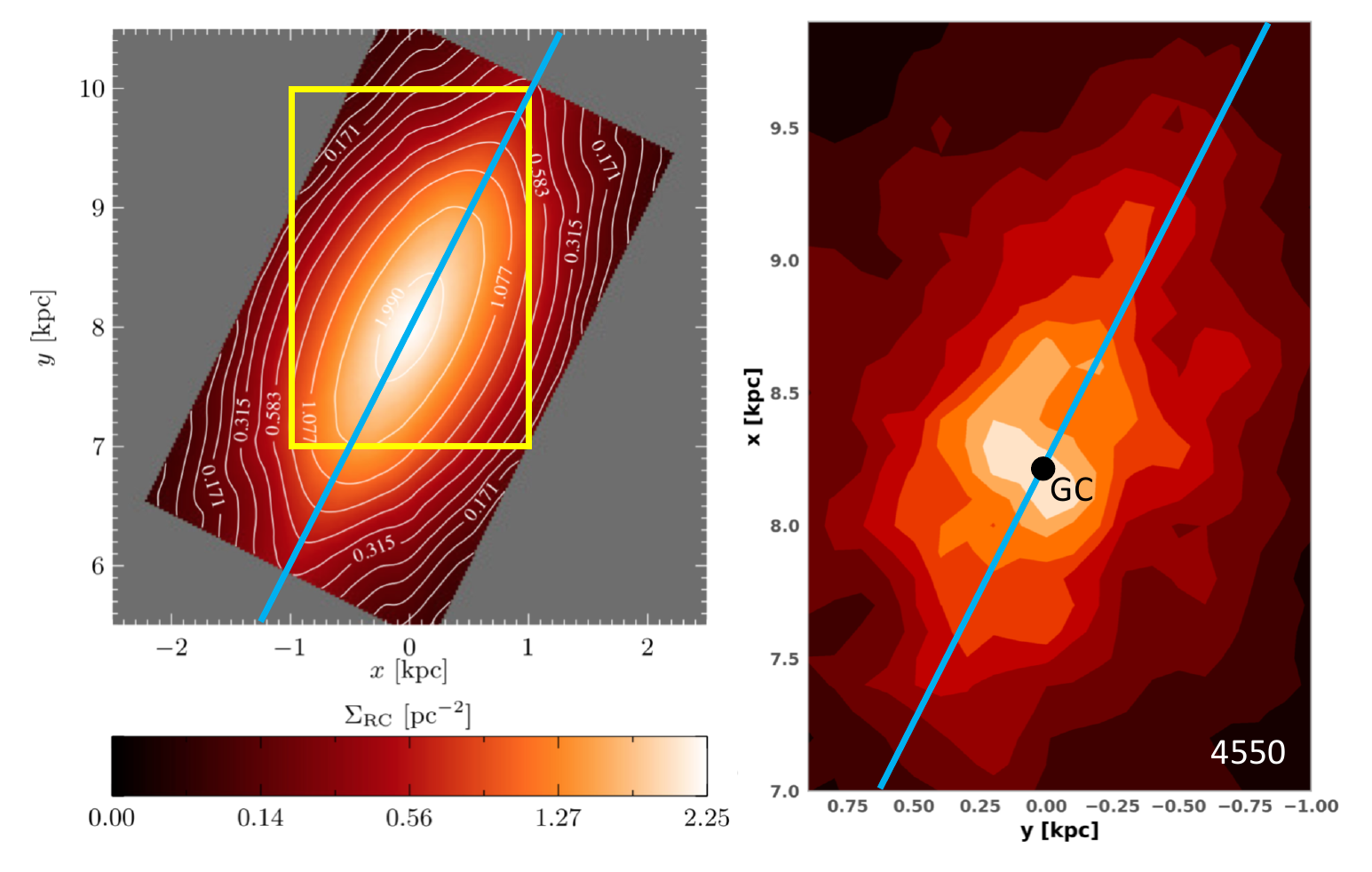}
 \caption{Surface density of RCGs (left panel, from \citealt[their Figure~17]{2013MNRAS.435.1874W}) and RRL stars (right panel). The yellow rectangle in the left panel shows the region covered by the right panel. The blue lines in both panels show an angle $\sim 27^\circ$, the RCG bar angle obtained from VVV \citep{2013MNRAS.435.1874W}. The right panel shows that this value is a reasonable approximation of the bar angle of our RRL distribution. The number in white in the bottom right corner of the right panel is the total number of the stars used here. The location of the Galactic centre is given by a black filled circle (GC).}
 \label{fig:up_side_down}
\end{figure*}

There has been a debate on whether bulge RRLs trace the bar structure. The bar angle determined using near-IR VVV observations of OGLE-III bulge RRLs by \citet{2013ApJ...776L..19D} is 12.5$^\circ\pm 0.5^\circ$, while optical observations of OGLE-IV bulge RRLs by \citet{2015ApJ...811..113P} gave a bar angle $20^\circ\pm 3^\circ$. We will use our sample to shed further light on this point. For the spatial distribution, we need not use Gaia data, so we use the dataset (15,599 sources) uncleaned with the Gaia RUWE flag to maintain the completeness of the RRab stars.

We plot the distribution of bulge RRLs from a top-down view. The left panel in Figure~\ref{fig:up_side_down} is from \citet[their Figure 17]{2013MNRAS.435.1874W} showing the surface density obtained for RCGs from the VVV survey; the right panel shows the surface density map from our data with the following operations:

\begin{itemize}

 \item We restrict the region to $-6.5^\circ<b<-2.8^\circ$, $-1\,{\rm kpc}<y<1\,{\rm kpc} , 7\,{\rm kpc}<x<10\,{\rm kpc}$ to avoid the incomplete fields of the survey.
 
 \item We segment the region into square cells of the size of $ 0.1$\,kpc from the top-down view. The color of each cell is determined by the star counts in a circle with a radius of $0.15$\,kpc centred in the cell's centre. The star counting region for each cell is larger than the cell itself for better statistics. This operation is similar to \textit{smoothing} in image reduction.
 
 \item Note that we consider the cone effect, i.e., multiply the count in each cell by a factor 1/$s^2$ where $s$ is the star-Sun distance.
 
\end{itemize}

We find a bar angle ($\alpha$) for RRLs $\sim$ 27$^\circ$, which is close to earlier determinations for younger populations, like VVV Type II Cepheids ($\sim 30^\circ$, \citealt{2019ApJ...883...58D}), Gaia bulge Miras ($\sim 21^\circ$, \citealt{2020MNRAS.492.3128G}), VVV RCGs ($20^\circ - 30^\circ$, \citealt{2017MNRAS.471.4323S}) , OGLE RCGs ($29^\circ \pm 2^\circ$, \citealt{2013MNRAS.434..595C}; $27^\circ\pm 2^\circ$,\citealt{2013MNRAS.435.1874W}; $25^\circ\pm 2^\circ$, \citealt{2007MNRAS.378.1064R}; $20^\circ - 30^\circ$, \citealt{1997ApJ...477..163S}), 2MASS red giant star counts ($20^\circ - 35^\circ$, \citealt{2005A&A...439..107L}), and  modeling the asymmetry of the COBE NIR photometry ($25^\circ \pm 10^\circ$; \citealt{1995ApJ...445..716D}, \citealt{1997MNRAS.288..365B}, \citealt{1999A&A...345..787F}, \citealt{2002MNRAS.330..591B}).

\begin{figure*}
 \includegraphics[width=0.8\linewidth]{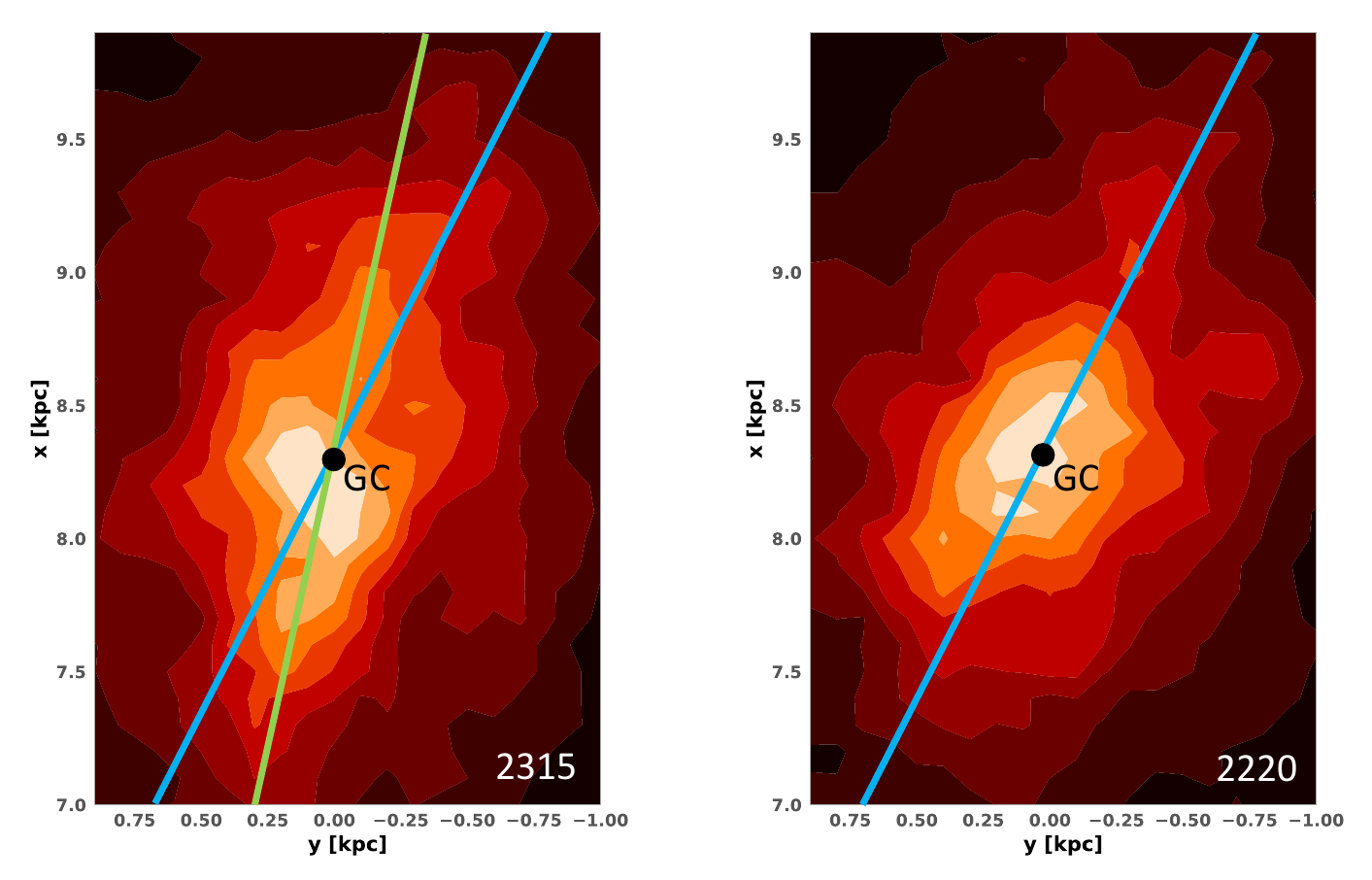}
 \caption{As the right panel of Figure~\ref{fig:up_side_down}, but after splitting our sample by
metallicity. The left panel is for metal-poor RRLs ([Fe/H]<$-1$), and the right panel is for metal-rich ones ([Fe/H]>$-1$). The blue lines show an angle $\sim 27^\circ$, and the green line shows a bar angle $\sim 12^\circ.5$, as given by \citet{2013ApJ...776L..19D}. We see that the bar angle for the metal-poor RRLs is smaller. Selection biases from different metallicity ranges can explain previous disagreements on the bar angle obtained from bulge RRLs (see section~\ref{sec:geo_dis}).}
 \label{fig:up_side_down_diff_met}
\end{figure*}

We next adopt the same instruction separately to the metal-poor ([Fe/H]$<-1$) and the metal-poor ([Fe/H]$>-1$) shown in Figure~\ref{fig:up_side_down_diff_met}. We see that the metal-poor RRLs show a smaller bar angle, which is in agreement with the bar angle measured by \citet{2013ApJ...776L..19D} and \citet{2017AJ....153..179M}, who did not cut the metallicity. So the discrepancy of the two bar angles of RRLs might be explained by the selection bias due to different metallicity ranges. We note that their sample is smaller, and may suffer from VVV photometry calibration issues \citep{2020ExA....49..217H}. Currently, their data are being re-analyzed by D\'ek\'any (2020, private communication).

We use the Gaussian resampling method mentioned at the beginning of section~\ref{sec:result} to confirm that structures in the star-count maps are not significantly influenced by the error of distances. We also used another distance dataset determined using period-luminosity-metallicity (PLZ) relationship from \citet{2015ApJ...808...50M} instead of \citet{2004ApJS..154..633C}, the distribution discrepancy of metal-poor and metal-rich RR Lyraes still holds.

\begin{figure}
 \includegraphics[width=\columnwidth]{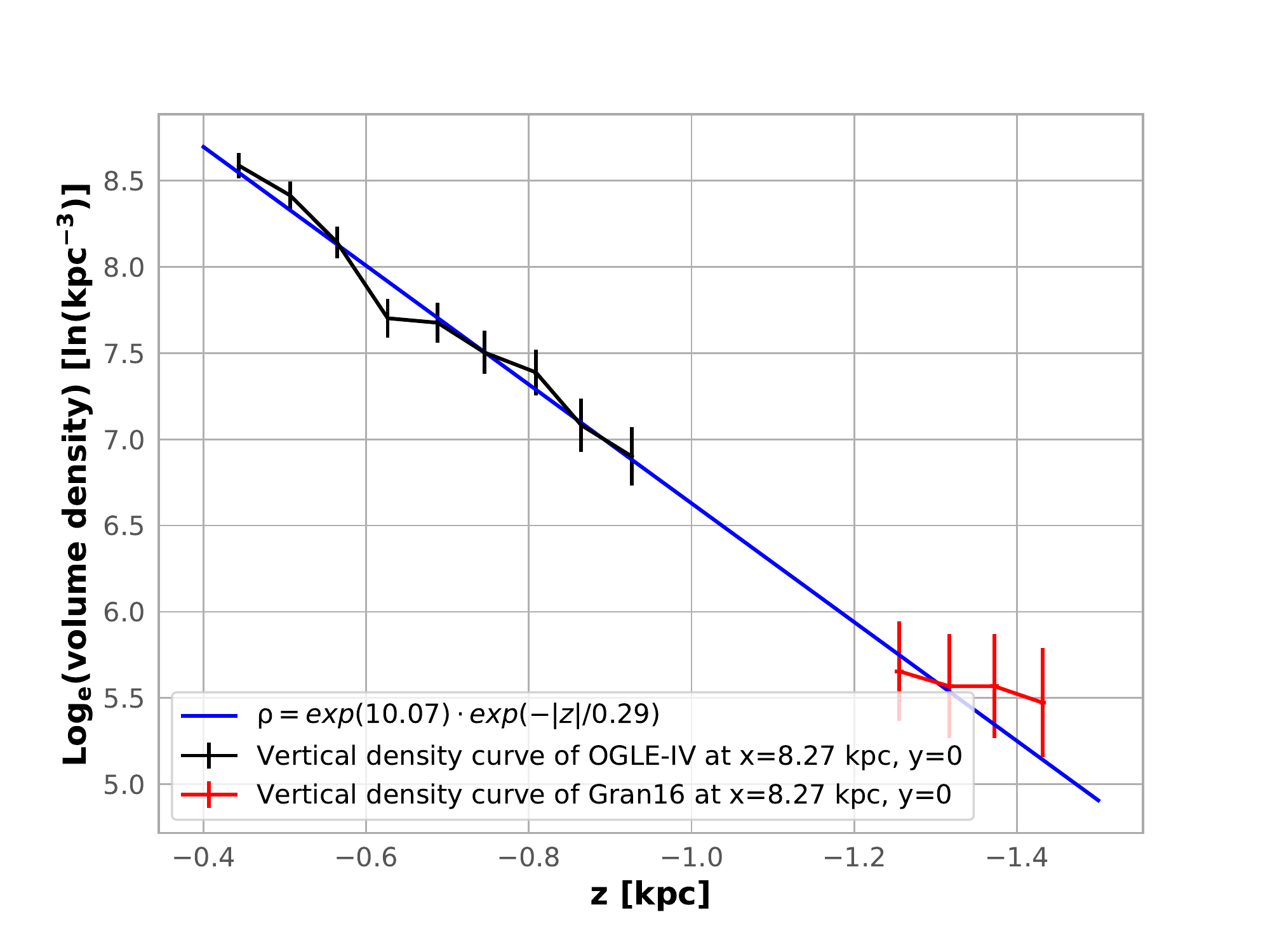}
 \caption{The vertical volume number density distribution in $\ln$ of the bulge RRLs. The vertical density of bulge RRLs follows an exponential distribution with a scale height $h_z=0.29\pm0.06$\,kpc. We note that the exponential scale length of the Galactic bar in $z$-direction traced by RC stars is $\sim 0.25$\,kpc \citep{2013MNRAS.434..595C}, indicating that the bulge RRLs follow similar vertical profile to the majority of the bar stars.}
 \label{fig:dens_curv}
\end{figure}

Figure~\ref{fig:dens_curv} shows the vertical volume number density distribution of the bulge RRLs at the Galactic centre. We calculate the density as follows: we select a rectangle field ($-4.2^\circ<l<4.2^\circ, -6.8^\circ<b<-2.8^\circ$) to ensure the field is all covered in our sample, then we get a cuboid which is in that field and is located at the Galactic centre:

\begin{equation}
\left\{
\begin{aligned}
x_{\rm 1}&<x<x_{\rm 2},\\
x_{\rm 1} \tan{l_{\rm 1}}&<y<x_{\rm 1}\tan{l_{\rm 2}},\\
x_{\rm 1}\tan{b_{\rm 1}}&<z<x_{\rm 2}\tan{b_{\rm 2}},
\end{aligned}
\right.
\end{equation}

\noindent where we use $x_1\equiv 8.0$\,kpc, $x_2\equiv 8.5$\,kpc, $l_1\equiv -4.2^\circ$, $l_2\equiv 4.2^\circ$, $b_1\equiv -6.8^\circ$, $b_2\equiv -2.8^\circ$. We cut $\delta z\equiv0.06$\,kpc in $z$-direction for each slice, for which we obtain the volume:

$$
V = (x_2-x_1)(y_2-y_1)\delta z.
$$
\noindent The volume number density is then obtained by $n = N/V$. The error bars are Poisson noise. We apply a similar workflow to the \citet{2016A&A...591A.145G} data. As shown in Figure~\ref{fig:dens_curv}, the vertical density of bulge RRLs follows an exponential distribution with a scale height $h_z=0.29\pm0.06$\,kpc. We note that the exponential scale length of the Galactic bar in $z$-direction traced by RC stars is $\sim 0.25$\,kpc \citep{2013MNRAS.434..595C}, indicating that the bulge RRLs follow the similar distribution of the majority of the bulge stars. The completeness of the RRab stars was discussed in \citet{2014AcA....64..177S, 2019AcA....69..321S, 2016A&A...591A.145G}, concluding that the RRab stars in these fields are sufficiently complete to do such statistics.

\subsection{Comparison of rotation curves}
\label{sec:comp_rot}

\begin{figure*}
 \includegraphics[width=\linewidth]{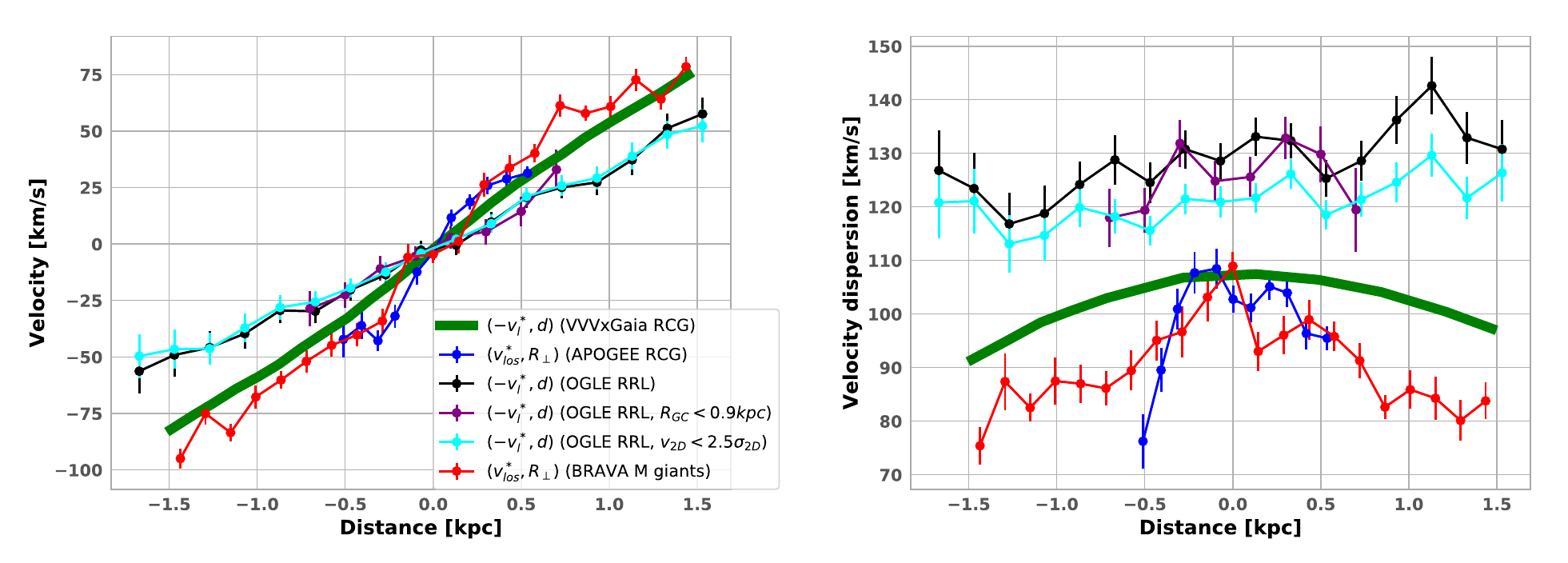}
 \caption{Left panel: Rotation curves from several surveys, the cross-match radius with Gaia DR2 is 0.2$\arcsec$ cleaned with RUWE<1.4. RRLs (black line) clearly rotate slower. The angular velocity (the slope of rotation curves fit in the inner 1.5\,kpc) derived from VVVxGaia \citep{2019MNRAS.487.5188S}, OGLExGaia and BRAVA \citep{2012AJ....143...57K} are respectively $(55.07\pm 0.43)$, $(36.95\pm 1.05)$, $(64.11\pm 5.43)$\,km\,s$^{-1}$\,kpc$^{-1}$ (see Table~\ref{tab:angular_velocity}). The consistency of the green line and the blue \& red lines is also a validation for the equivalence of these two indicators. Right panel: bulge RRLs are are kinematically hotter than other populations. Here the \textit{Distance} as $x$-axis is $R_\perp$ or $d$ as defined in Figure~\ref{fig:geo}. We then restrict RRLs to those within the inner 0.9\,kpc (purple line), or exclude high-velocity stars (cyan line) to check their influence (to compare with \citealt{2020AJ....159..270K}). We find that the rotation trend does not change significantly.}
 \label{fig:surveys}
\end{figure*}

\begin{table}
 \caption{Angular velocities derived from rotation curves.}
 \label{tab:angular_velocity}
 \begin{tabular}{lccc}
  \hline
  Survey & Population & Indicator & Ang. Vel. [km\,s$^{-1}$\,kpc$^{-1}$]\\
  \hline
  OGLExGaia & RRLs & ($-v_l^*,d$) & $36.95\pm 1.05$\\
  VVVxGaia & RCGs & ($-v_l^*,d$) & $55.07\pm 0.43$\\
  BRAVA & M giants & ($v_{\rm los},R_\perp$) & $64.11\pm 5.43$\\
  \hline
 \end{tabular}
\end{table}

In Figure~\ref{fig:surveys}, we apply the equivalence of indicators described in section~\ref{sec:comp} to the surveys from Table~\ref{tab:surveys}. Given the cylindrical rotation nature of RRLs, we can combine RRLs from different Galactic latitudes together. 

As mentioned in section~\ref{sec:vali_real}, the consistency (see Figure~\ref{fig:surveys}) of VVVxGaia (green line) and BRAVA \& APOGEE (blue \& red line) is also a validation for the equivalence of the two indicators (-$v_l^*,d$) and ($v_{\rm los}, R_\perp$): they describe similar populations and show similar properties.

RRLs show a significantly different behavior from RCG and M giants to which we compare them in Figure~\ref{fig:surveys}. This is presumably due to the fact that they are a
different population with slower rotation and are kinematically hotter, as found previously \citep{2016ApJ...821L..25K, 2018ApJ...863...79C}. The angular velocities derived from the rotation curves are shown in Table~\ref{tab:angular_velocity}. The angular velocity of RRLs (35\,km\,s$^{-1}$\,kpc$^{-1}$) is consistent with \citet{2019MNRAS.485.3296W} in which the rotation speed of RRLs at the distance of 1.5 kpc from the Galactic centre is $\sim$ 50 km\,s$^{-1}$. Also, it is worth mentioning here the fact that the metal-poor bulge stars rotate slower has been observed by several previous surveys \citep{2013MNRAS.432.2092N, 2016ApJ...819....2N,2017A&A...599A..12Z,2017A&A...601A.140R,2018ApJ...858...46C,2020MNRAS.491L..11A}. \citet{2013MNRAS.432.2092N} mentioned that the slower rotation in their metal-poor sample may be caused by stars of the metal-weak thick disc and halo which presently lie in the inner Galaxy, which might also explain the behaviour of bulge RRLs.

We then restrict the RRLs to those within the innermost 0.9\,kpc in order to compare with \citet{2020AJ....159..270K}; we take the value 0.9\,kpc from their Figure~6, which they treated as a central/classical bulge region. In our sample, the trend of RRLs does not change significantly with this constraint, and the velocity dispersion with such constraint is still significantly larger than younger populations. The difference between the two is that in \citet{2020AJ....159..270K}, stars with apocenter distances > 3.5\,kpc were not included; the exclusion of these stars also decreases the velocity dispersion reported in the \citet{2020AJ....159..270K} RRL dispersion as compared to what is found here. We will return to this issue briefly in the conclusion section.

\section{Discussion}
\label{sec:discussion}

\subsection{The equivalence of PM and LOS velocities from the dynamics point of view}
\label{sec:eq_pm_los}

A further point that needs to be discussed here is our finding that PM and LOS velocities can be indiscriminately used to study the rotation in the bulge region of our Galaxy. This can be shown (see section~\ref{sec:geo}) in a simple, geometric way under the following two assumptions: The system is in equilibrium, the backbone, main periodic orbits in the region under consideration are not far from circular, and the rotation in the inner parts increases linearly with distance from the centre. The first and third of these assumptions should not present any difficulties. The second one, however, namely that the orbits can be described by circles, needs further discussion as it is by now well established that our Galaxy is barred, and we are modeling the inner regions. For this reason, we used mock data from two independent simulations, which we chose to be as different as possible (see section~\ref{sec:2sim}). Although the snapshots we chose are barred, we found that, contrary to what one would have naively expected, our two rotation indicators can be indiscriminately used in both models.

To understand this, one must remember that the region we are studying here is not the whole bar, but only its centre-most part, i.e., stars which are at a distance of $1-2$ kpc only from the centre, or, equivalently, less or of the order of half the bar length. At such distances, the amplitude of the $m=2$ and of the higher even Fourier components ($m=4, 6, 8,...$) of the surface density distribution, are much lower than somewhat further out, as has been shown both by observations (e.g., \citealt{2006AJ....132.1859B}) and simulations (e.g., \citealt{2002MNRAS.330...35A}). This argument can be pushed further by looking at the shape of the main periodic orbits in that region. Indeed, \citet[their Figure~9]{1992MNRAS.259..328A} gives information on the shape of these periodic orbits\footnote{This work concerns the two main families of periodic orbits at the centre, namely the x1 and the x2,  therefore, also the orbits trapped around them. Such information does not yet exist for higher multiplicity periodic orbits, but their importance in the dynamics of this inner region is also not yet established.} and shows that their elongation is relatively small near the centre, then increases with increasing distance from the centre to reach a maximum and then drops in the outer parts of the bar. Of course, the axial ratio of these orbits will depend on the gravitational potential and particularly on the shape of the iso-potentials in the central regions, so that it is not possible, from these orbits only, to make any quantitative statements for the inner part of the Galactic bar. Qualitatively, however, our arguments here show that the rotation obtained by PM and LOS velocities can be equivalent, even when the bar is present.

\subsection{Why was the rotation of bulge RRLs not previously observed quantitatively?}
\label{sec:non_rot}

\begin{figure*}
 \includegraphics[width=\linewidth]{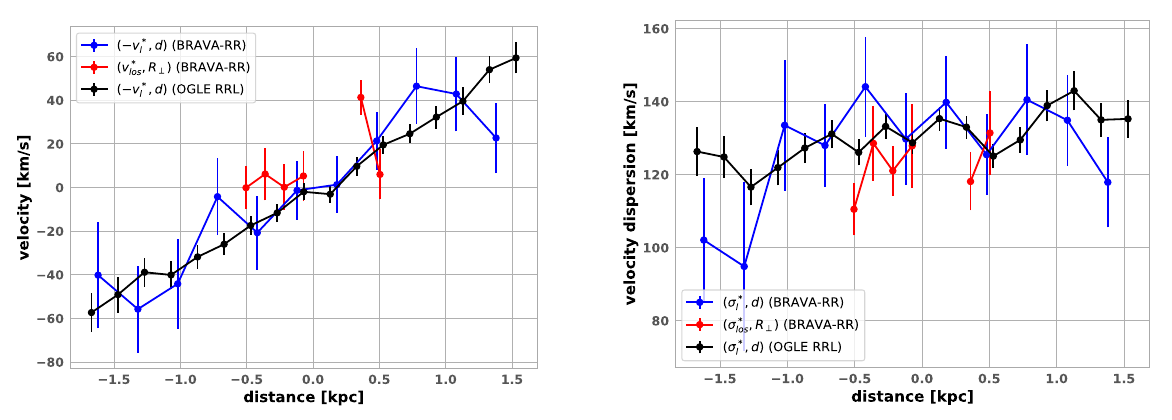}
 \caption{We show why \citet{2016ApJ...821L..25K} have not observed the rotation. Their sample is too close to the Galactic centre ($l$ range is too small) so that $v_{\rm los}$ is not a good indicator to reflect the rotation. The blue line is the 6D sample obtained after cross-matching BRAVA-RR and Gaia DR2, showing consistency with both BRAVA-RR and ours (black line).}
 \label{fig:brava_rr}
\end{figure*}

Independent observations from BRAVA-RR \citep{2016ApJ...821L..25K} and VVV \citep{2018ApJ...863...79C} show that bulge RRLs are non-rotating or slow-rotating. We demonstrate that this is qualitatively, but not quantitatively, consistent with our results.

The BRAVA-RR presented by \citet{2016ApJ...821L..25K} is an extension program of BRAVA, which focused on the RRLs in the Galactic bulge with LOS velocities. The sample of 947 sources was selected from OGLE-III RRL catalog, whose field is shown in \citet[their Figure 1]{2016ApJ...821L..25K} ranging  $-5^\circ\lesssim l\lesssim 5^\circ, -5.5^\circ \lesssim b\lesssim -1^\circ$. Figure~\ref{fig:brava_rr} is prepared as follows: after cross-matching BRAVA-RR with Gaia DR2, we obtain a sample of 862 RRLs with 6D phase space information. Then we can use both indicators ($v_{\rm los}^*, R_\perp$) and ($-v_l^*, d$) to obtain the angular velocity. We see, within error bars, the blue line is consistent with the other two curves. Because the BRAVA-RR curve (red) is too close to the Galactic centre, $(v_{\rm los}^*, R_\perp)$ is not a good indicator to reflect the rotation. In the right panel of Figure~\ref{fig:brava_rr}, we plot the corresponding velocity dispersion curves and show again that the three curves are consistent with each other.

\begin{figure}
 \includegraphics[width=\columnwidth]{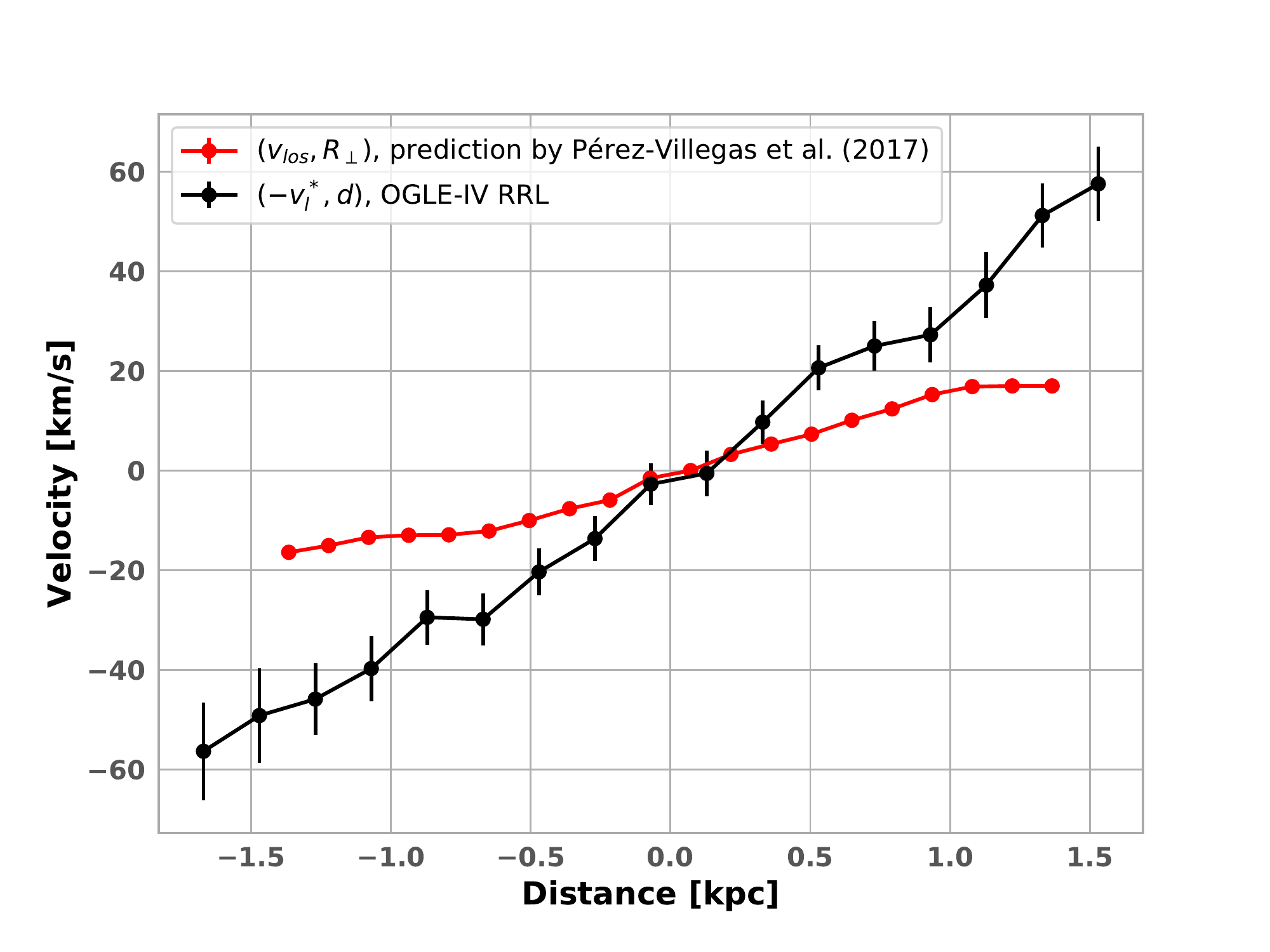}
 \caption{The simulation by \citet{2017MNRAS.464L..80P} reproduced the BRAVA-RR \citep{2016ApJ...821L..25K} data. The assumption of \citet{2017MNRAS.464L..80P} is that RRLs in the Galactic bulge are inner halo stars; the predicted rotation is too slow to be in agreement with the rotation curve based on Gaia proper motions.}
 \label{fig:perez}
\end{figure}


Figure~\ref{fig:perez} shows the rotation curve for the inner stellar halo, as predicted by a simulation of \citet{2017MNRAS.464L..80P} and obtained from the Gaia proper motions. We extracted the values from the bottom left panel of their Figure~3. We also added our result from Figure~\ref{fig:surveys}. We see there is a considerable difference between the two, which could well indicate that the RRL population of the Galactic bulge is not necessarily consistent with being the inward extension of the Galactic metal-poor stellar halo. We note that the inconsistency between the simulation in \citet{2017MNRAS.464L..80P} and Gaia observation was also mentioned by \citet{2019MNRAS.485.3296W}.

Another piece of evidence stating their non-rotating RRL paradigm is provided by \citet[their Figure 13]{2018ApJ...863...79C}, based on VVV data. Their RRL identification is made by an automated RRab classifier based on machine learning \citep{2016A&A...595A..82E} and its modified version \citep[their appendix]{2018ApJ...863...79C}. Their sample of 959 RRLs in the field $|l|,|b|\lesssim 1.7^\circ$ is too close to the disk to be cross-matched with Gaia DR2. 

\begin{figure}
 \includegraphics[width=\columnwidth]{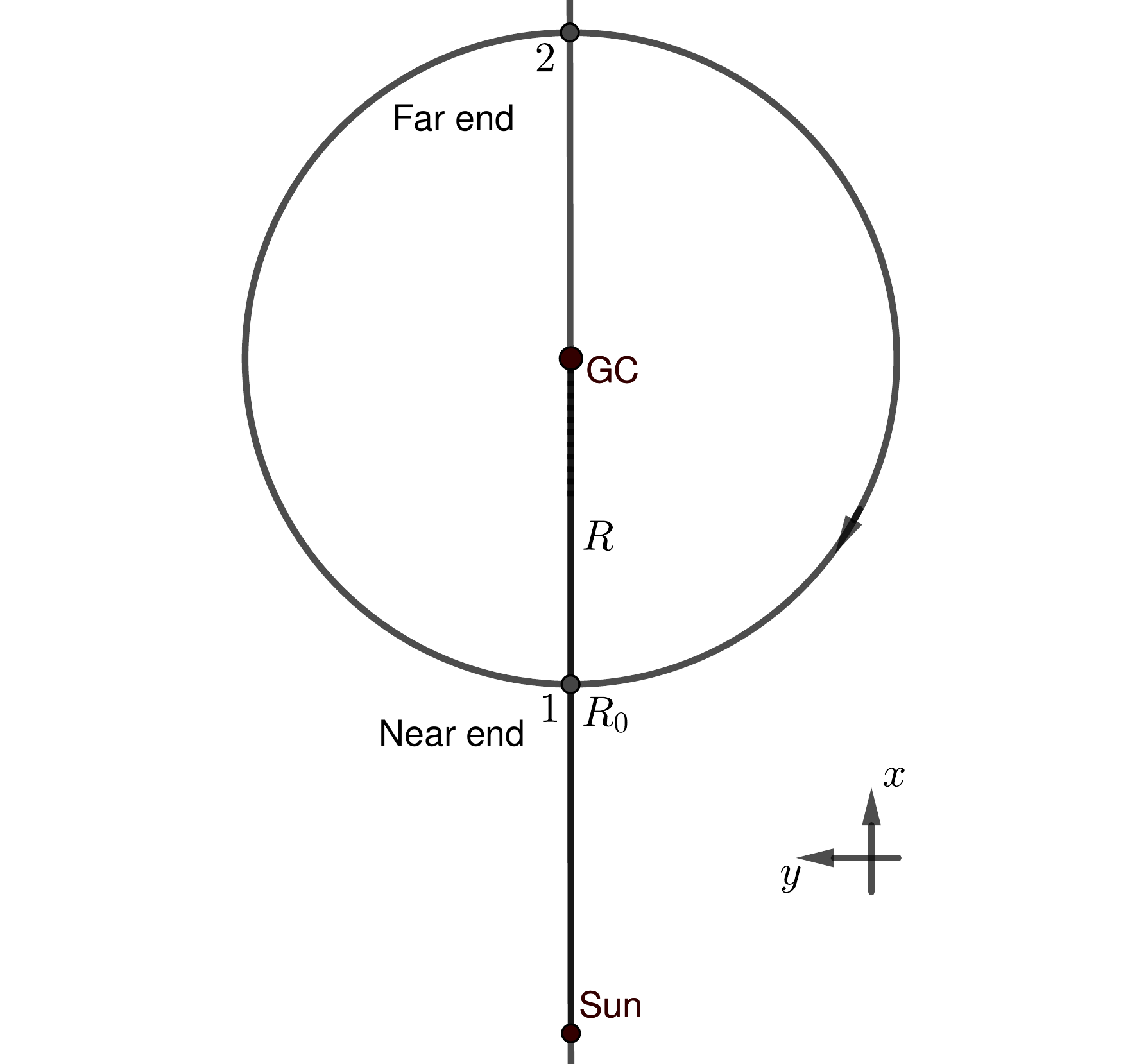}
 \caption{A simplest view from the North of Galactic plane to understand the trend in the $\mu_{l}$-distance diagrams, so that to explain why \citet{2018ApJ...863...79C} did not find the rotation. }
 \label{fig:ramos_geo}
\end{figure}

As shown in \citet[their Figure 13]{2018ApJ...863...79C}, they use $\mu_l^*$ as the $y$-axis and $W_{K_s}$-band magnitude as the $x$-axis, fitting the sample of stars with a horizontal line. However, this does not support their non-rotating interpretation but supports our scenario. 

To understand this better, firstly we take the observables into Equation~\ref{eq:v_l}. Of the solar motion (see Figure~\ref{fig:geo}), we know that:
\begin{equation}
\begin{aligned}
v_{\odot,y}& \equiv |\mu_{\rm GC}|\cdot R_0,\\
\mu_l\cdot s&=v_{l}-v_{\rm \odot,T}=v_{l}-v_{\odot,y}\cos{l},
\end{aligned}
\end{equation}

\noindent where $v_{\odot,y}$ is the solar tangential velocity around the Galactic centre and $|\mu_{\rm GC}|=30.24$\,km\,s$^{-1}$\,kpc$^{-1}$=$-\mu_{\rm GC}$ is the proper motion of the Galactic centre, $v_{\rm \odot, T}$ is the tangential velocity with respect to the line of sight (see Figure~\ref{fig:geo}). We have: 
\begin{equation}
\begin{aligned}
 v_{l}=\mu_l\cdot s + |\mu_{\rm GC}|\cdot R_{0}\cos{l}.
\end{aligned}
\end{equation}

Then we take the simplest case into consideration (see Figure~\ref{fig:ramos_geo}, see also  \citealt[their section~3.5]{2015ApJ...808...75Q}) for a qualitative conclusion. When we look along the Sun-GC line on the Equatorial Plane ($l=0, b=0, \alpha=0  \,{\rm or}\, 180^\circ$), For stars on pure circular orbits, the PM of the stars in the near-end and far-end are:

\begin{equation}
\begin{aligned}
\mu_{l,1}=\frac{\mu_{\rm GC}\cdot R_0 +\omega\cdot R}{R_0-R}\\
\mu_{l,2}=\frac{\mu_{\rm GC}\cdot R_0 -\omega\cdot R}{R_0+R},
\end{aligned}
\end{equation}

\noindent where $\mu_{GC}$=-6.38~mas/yr=-30.24\,km\,s$^{-1}$\,kpc$^{-1}$. Here we can see, for a rigid body rotation with angular velocity $\omega$, that there will be three cases:

\begin{itemize}
 \item In the case of $\omega=|\mu_{\rm GC}|=-\mu_{\rm GC}$, we will obtain $\mu_{l,1}=\mu_{l,2}=\mu_{\rm GC}=-6.38$~mas/yr, or a horizontal line in the $\mu_l$-distance diagram.
 
  \item For $\omega>|\mu_{\rm GC}|$, $\mu_{l,1}>\mu_{\rm GC}>\mu_{l,2}$, we obtain a descending trend in the $\mu_l$-distance diagram.

  \item If $\omega<|\mu_{\rm GC}|$, $\mu_{l,1}<\mu_{\rm GC}<\mu_{l,2}$, we obtain an ascending trend in the $\mu_l$-distance diagram.
\end{itemize}
 
\begin{figure}
 \includegraphics[width=\columnwidth]{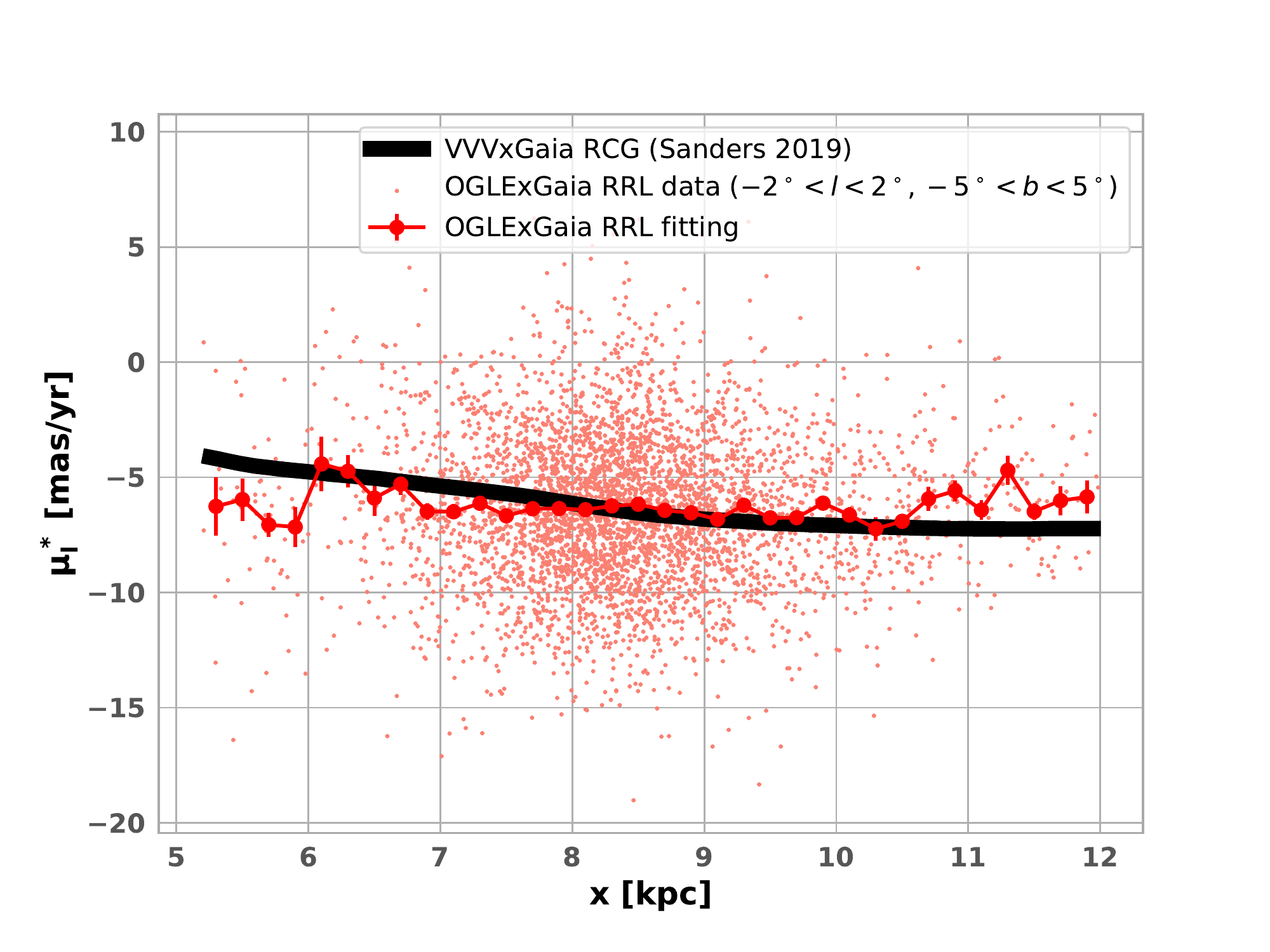}
 \caption{Proper motions of RRLs and RCGs in the bulge. We compare this to \citet[their Figure 13]{2018ApJ...863...79C}; the horizontal trend in the bulge region (7$\sim$10\,kpc) reflect a rotating sample with similar angular velocity to the Sun's around the Galactic centre.}
 \label{fig:rot_ramos}
\end{figure}

Our angular velocity of bulge RRLs ($36.95\pm 1.05$)\,km\,s$^{-1}$\,kpc$^{-1}$ is slightly larger than that of $|\mu_{\rm GC}|$, which is close to the first case. We can also check this with our data in Figure~\ref{fig:rot_ramos}, in which the horizontal trend in the bulge region (7$\sim$10\,kpc from the Sun) reflects a rotating sample with similar angular velocity to the Sun's rotation around the Galactic centre.

Moreover, as similarly explained in section~\ref{sec:spatial}, the fact that the rotation of bulge RRLs was not clearly observed might also be explained by the selection bias on metallicity; extremely metal-poor RRLs show mild rotation.

\section{Conclusion}
\label{sec:conclusion}

Spectroscopic surveys of the Galactic bulge region show that this component has a complex chemodynamical structure \citep[and references therein]{2018ARA&A..56..223B}. In our work, "bulge" is a general concept referring to the specific Galactic centre region (sometimes referred to as the "inner Galaxy"). We provide a phenomenological view of the RRLs in the inner Galaxy region for a better understanding of their nature.

The proper motions have been, so far, rarely used to obtain the rotation curve in the Galactic bulge due to their measurement difficulty. Thanks to the data release 2 of Gaia and the OGLE-IV RRLs with distance determination, we can now use PM to demonstrate the bulge rotation and compare it with previous LOS velocities. We summarize our results as follows:

\begin{itemize}

 \item We show the equivalence of PM and LOS velocities as the Galactic bulge rotation indicators, which has not been discussed in previous works.
 
 \item Metal-poor and metal-rich RRLs show different behaviors: 1) their angular velocities are different. As seen in Figure~\ref{fig:rot_diff_met}, this is $(32.42\pm 1.48)$\,km\,s$^{-1}$\,kpc$^{-1}$ for [Fe/H]<$-1$ and  $(40.29\pm 2.28)$\,km\,s$^{-1}$\,kpc$^{-1}$ for [Fe/H]>$-1$; 2) their spatial distributions are different, the metal-rich RRLs show a similar triaxial structure to that of RCGs while the metal-poor ones show a smaller bar angle. This provides a view to explain the conflict between \citet{2013ApJ...776L..19D} and \citet{2015ApJ...811..113P}; 3) extremely metal-rich RRLs ([Fe/H]>$-0.5$) show similar behavior as RCGs.

 \item We update the results of \citet{2016ApJ...821L..25K} and \citet{2018ApJ...863...79C} quantitatively: bulge RRLs have an angular velocity of about $35$\,km\,s$^{-1}$\,kpc$^{-1}$, i.e., are considerably slower than younger bulge stars (50-60\,km\,s$^{-1}$\,kpc$^{-1}$). This result is consistent with \citet{2019MNRAS.485.3296W}, whose sample of halo RRLs rotates with 50\,km\,s$^{-1}$ in their innermost radial bin (1.5\,kpc to the Galactic centre). We also find the absence of the velocity dispersion central peak for bulge RRLs with $|b|<3^\circ$, which is similar to, but more extreme than that for metal-poor RCGs \citep[their Figure~6]{2013MNRAS.432.2092N}.
 
\end{itemize}

 The halo contamination is not easy to determine, but we offer some clues about the halo/classical bulge fraction (see also  \citealt{2016ApJ...821L..25K, 2017MNRAS.464L..80P}). We note that \citet{2013MNRAS.432.2092N} mentioned that the slower rotation in their metal-poor sample may be caused by stars of the metal-weak thick disc and halo which presently lie in the inner Galaxy. This interpretation could also apply to the bulge RRLs. Firstly, the vertical number density of bulge RRLs is distributed exponentially (see Figure~\ref{fig:dens_curv}), not necessarily obeying the stellar-halo-like power law; secondly, we back the claim of \citet{2019MNRAS.485.3296W} that the rotation predicted by the \citet{2017MNRAS.464L..80P} model is too low to be in agreement with the observed rotation from the Gaia DR2 proper motions (see Figure~\ref{fig:perez}), which moderates the claim of \citet{2017MNRAS.464L..80P} that the RR Lyrae population in the Galactic bulge is consistent with being the inward extension of the Galactic metal-poor stellar halo; thirdly, \citet[a preprint which appeared when our paper was being refereed]{2020arXiv200612507S} analysed the chemodynamical nature of bulge RRLs based on \citet{2016ApJ...821L..25K} data, and found "the existence of a breaking point in the halo properties at around 5\,kpc", which they described as "open to the possibility that the innermost stellar halo is somehow different from its large-scale counterpart". 
 



While the paper was being refereed, a preprint by \citet{2020AJ....159..270K} appeared on astro-ph. They presented the results from BRAVA-RR DR2 (cross matched with Gaia DR2), which is a LOS-velocity survey toward RRLs in the Galactic bulge. Their sample size is a factor of 5.9 smaller than ours, but has a substantial advantage since they have 6D phase-space information. They defined "Halo RRL" as the stars with apocenter distances > 3.5\,kpc, which composes about 25\% of stars in their sample \footnote{Another recent work \citep{2020arXiv200705849P}, which appeared when our work was being refereed, also concluded that $\sim$25\% of the central bulge may be from the halo.}. Since we have not traced the orbits of stars (which is model dependent), we cannot determine this halo fraction precisely. They also found a substantial fraction of "high-velocity stars" ($\sim$7\% of stars have 3D velocities greater than 2.5$\sigma$ from the mean of the distribution); in our sample, the fraction of stars with 2D velocities higher than 2.5$\sigma_{\rm 2D}$=380\,km\,s$^{-1}$ is about 2\%, a factor of 3.5 lower than their fraction. It is unclear whether this can be well explained by the difference between 2D and 3D velocities. Also, their Figure~8 showed that the "Central/Classical RRL" (with RGC < 0.9\,kpc) have, as mentioned in section~\ref{sec:comp_rot}, velocity dispersion similar to younger populations. Furthermore, their sample are non- or slowly rotating (different from "Bar/Bulge RRL"). This is different from the rotation inferred from Gaia proper motions (Figure~\ref{fig:surveys}). The difference is likely due to two reasons: 1) our samples are different since they have decontaminated "Halo RRL" (stars with apocenter distances > 3.5\,kpc as they defined) while we did not; 2) their sample size is smaller which leads to somewhat larger error bars and make their rotation signal more difficult to see. We are currently studying the issue with radial velocities from APOGEE DR16 to further understand the kinematics of the RRLs in the inner Galaxy. We will present our results in a future paper.

\section*{Acknowledgements}

We thank the referee for insightful comments that improved the paper. We thank Istv\'an D\'ek\'any and Marina Rejkuba for communication on the extinction issue. The research presented here is partially supported by the National Key R\&D Program of China under grant No. 2018YFA0404501, by the National Natural Science Foundation of China under grant Nos. 11773052, 11761131016, 11333003, and by a China-Chile joint grant from CASSACA. E.A. thanks the CNES for
financial support. This work was granted access to the HPC resources of CINES under the allocations 2018-A0040407665 and 2019-A0040407665 attributed by GENCI (Grand Equipement National de Calcul Intensif). Center de Calcul Intensif Aix-Marseille is acknowledged for granting access to its high-performance computing resources. These two CPU time allocations allowed EA to run and analyze the gas-rich simulation used here. J.S. acknowledges support from a {\it Newton Advanced Fellowship} awarded by the Royal Society and the Newton Fund. This work made use of the facilities of the Center for High Performance Computing at Shanghai Astronomical Observatory. P.P. has been supported by the National Science Centre,
Poland, grant OPUS 2016/23/B/ST9/00655. The original idea was boosted in the 2018 Gaia-LAMOST Sprint workshop, supported by the NSFC under grants 11333003 and 11390372. This work has made use of data from the European Space Agency (ESA) mission {\it Gaia} (\url{https://www.cosmos.esa.int/gaia}), processed by the {\it Gaia} Data Processing and Analysis Consortium (DPAC, \url{https://www.cosmos.esa.int/web/gaia/dpac/consortium}). Funding for the DPAC has been provided by national institutions, in particular the institutions participating in the {\it Gaia} Multilateral Agreement.

\section*{Data availability}
The data generated as part of this project may be shared on a reasonable request to the corresponding author.

\label{lastpage}
\end{document}